\documentclass[a4paper,11pt]{article}
\usepackage{jheppub} 
\usepackage{lineno}
\usepackage{braket}
\usepackage{mathtools}
\usepackage{bbm}
\usepackage{lipsum}
\usepackage{tikz}
\usepackage{tikz-cd}
\usepackage{quantikz}
\usepackage{graphicx}
\usepackage{subcaption}
\usepackage{amsmath}
\usepackage{lipsum}
\usetikzlibrary{quantikz2,positioning, fit, backgrounds}

\usepackage{booktabs}

\usepackage{amssymb}   
\usepackage{bm}    

\usepackage{amsthm}
\usepackage{comment}

\usepackage[dvipsnames]{xcolor}
\usepackage{placeins}
\usepackage[normalem]{ulem} 

\newtheoremstyle{nopunct}%
  {3pt}
  {3pt}
  {\itshape}
  {}
  {\bfseries}
  {}
  { }
  {}

\theoremstyle{nopunct}

\usepackage{soul}
\setstcolor{red}

\title{Quantum Graph Neural Networks for Jet Tagging on Quantum Hardware}

\author[a]{Benjamin Jobilal,}
\author[b,c]{Jinghong Yang,}
\author[b]{Trevor Smith,}
\author[b]{Vincent Calvo,}
\author[a, d, e]{Zhong-Bo Kang,}
\author[b,c]{Shabnam Jabeen}

\affiliation[a]{Department of Physics and Astronomy, University of California, Los Angeles, CA 90095, U.S.A.}
\affiliation[b]{Department of Physics, University of Maryland, College Park, MD 20742, U.S.A.}
\affiliation[c]{National Quantum Lab, University of Maryland, College Park, MD 20742, U.S.A.}
\affiliation[d]{Mani L. Bhaumik Institute for Theoretical Physics, University of California, Los Angeles, CA 90095, U.S.A.}

\affiliation[e]{Center for Quantum Science and Engineering, University of California, Los Angeles, CA 90095, U.S.A}

\emailAdd{bjobilal@g.ucla.edu}
\emailAdd{yangjh@umd.edu}
\emailAdd{zkang@physics.ucla.edu}
\emailAdd{jabeen@umd.edu}

\abstract{Jets are central to the physics programs of both current and future colliders, from precision Standard Model measurements and searches for new physics at the Large Hadron Collider to studies of nucleon structure at the future Electron-Ion Collider. Motivated by these applications, we explore quantum machine learning for jet classification and present a permutation-invariant Quantum Graph Neural Network (QGNN) applied to particle-cloud representations of jets. We apply the model to two such discrimination tasks: quark vs. gluon and up vs. down quark flavor tagging, with the latter being, to our knowledge, the first application of a quantum model to this problem. In the ideal simulation, the QGNN performs competitively against the Particle Flow Network and traditional QCD observables. We further deploy scaled-down models to IBM and IonQ quantum processing units (QPUs), where we train and evaluate them, obtaining promising results. Finally, we perform an interpretability analysis to characterize the observables learned by the quantum model, relating them to generalized angularities for the quark-gluon study and to jet charge for the flavor study. 
}

\begin{document}
\maketitle

\section{Introduction}

Jets are highly energetic, collimated sprays of particles produced in high-energy particle collisions. At colliders such as the Large Hadron Collider (LHC), complex Quantum Chromodynamics (QCD) processes produce energetic quarks and gluons that initiate parton showers and subsequently hadronize into jets. Jets are therefore of central importance to precision measurements and searches for physics beyond the Standard Model~\cite{Larkoski:2017jix}. For example, quark-gluon discrimination can improve the separation of rare signals from large QCD backgrounds, including searches involving specific Higgs production and decay modes~\cite{FerreiradeLima:2016gcz, Cho:2020ftg}. Discrimination among quark flavors provides sensitivity to flavor-dependent Parton Distribution Functions (PDFs)~\cite{Arratia:2020azl} and supports measurements of electroweak and top-quark properties, such as $W$-boson polarization and top-quark couplings~\cite{ATLAS:2016fbc, CMS:2020ezf, Subba:2022czw}. Flavor discrimination is particularly important for the spin-physics program at the Electron-Ion Collider (EIC)~\cite{Accardi:2012qut}, where different quark flavors can contribute with different signs to spin-dependent observables. Their contributions may therefore partially cancel in flavor-inclusive measurements, making flavor separation essential for resolving the underlying flavor-dependent spin structure of the nucleon~\cite{AbdulKhalek:2021gbh}. More broadly, such flavor-tagging capabilities will also be valuable at future facilities such as the FCC-ee. Reliable jet classification is therefore an important component of experimental high-energy physics.

In pursuit of improved jet classification, a wide range of approaches have been developed, ranging from physics-motivated jet observables to modern machine learning algorithms~\cite{Larkoski:2017jix}. Traditional observables are computed from the particles comprising a jet and are designed to capture specific features of its substructure. Examples include jet charge, which is commonly used to distinguish jets initiated by up and down quarks~\cite{Field:1977fa,Fraser:2018ieu,Kang:2020fka,Kang:2021ryr}, jet girth~\cite{Almeida:2008yp, Yan:2020zrz}, and $N$-subjettiness, which probes the degree to which a jet has $N$ subjets~\cite{Thaler:2010tr}. While such observables are often physically interpretable and theoretically well motivated, they are ultimately limited by the information they are explicitly constructed to encode. In recent years, the field has increasingly shifted toward machine learning approaches, driven by the availability of large simulated datasets and the ability of neural networks to learn complex, high-dimensional correlations directly from low-level particle information. Some early approaches treated jets as images in the $\eta$--$\phi$ plane and applied convolutional neural networks to identify patterns in the radiation profile and pronged substructure of the jet~\cite{Cogan:2014oua, Andrews:2021ejw}. More recently, permutation-invariant architectures include graph neural networks such as ParticleNet~\cite{Qu:2019gqs}, transformer-based models such as ParT~\cite{Qu:2022mxj}, and deep-set architectures such as the Particle Flow Network (PFN) and Energy Flow Network (EFN)~\cite{Komiske:2018cqr}. These have demonstrated state-of-the-art performance on a variety of jet-tagging tasks. Collectively, these methods have significantly advanced jet classification by leveraging information that is difficult to capture with closed-form observables alone, although their increasing complexity can come at the cost of interpretability~\cite{Wetzel:2025uhj}.

Quantum computing has emerged as a promising computational paradigm for tackling problems that may be challenging for classical computers. One important application is quantum simulation, where quantum processors can be used to study the dynamics of quantum systems, including gauge theories such as the Schwinger model~\cite{Zohar:2013zla, Lamm:2019bik, Kokail:2018eiw, Klco:2018kyo, Shaw:2020udc, Farrell:2023fgd, Farrell:2024fit, Davoudi:2024wyv, Ikeda:2025bjb}. Beyond quantum simulation, quantum computing can also be used as a platform for machine learning, giving rise to Quantum Machine Learning (QML)~\cite{Biamonte:2016ugo}. One widely used approach, which we adopt here, encodes classical data into a parametrized quantum circuit whose tunable parameters are optimized during training, with measurements of the circuit providing the model output for a given input. While QML has been explored in a variety of applications, its use for jet classification remains relatively nascent. We therefore view jet tagging on current quantum hardware as a useful setting for assessing the feasibility of quantum machine learning and for exploring the capabilities and limitations of QML in a relevant high-energy physics task.

The classical architectures described above have inspired corresponding quantum models, including Quantum Convolutional Neural Networks (QCNNs)~\cite{Elhag:2024xfw}, Quantum Graph Neural Networks (QGNNs)~\cite{Jahin:2024zss, Li:2026ydk, Jahin:2024wjw}, and other approaches~\cite{Bal:2025ydm}, which have been applied to jet classification tasks such as quark–gluon discrimination and have demonstrated competitive performance in idealized simulations. However, implementations on real quantum hardware remain scarce, largely because the limited scale and fidelity of current quantum processors constrain the depth and size of practical QML models. Across high-energy physics, direct training on quantum hardware has so far been demonstrated only in a small number of proof-of-principle, event-level studies~\cite{Wu:2022tnc, Terashi:2020wfi}; for jet tagging in particular, quantum hardware has been used only for inference, with models trained in classical simulation~\cite{Chen:2024rna, Napolitano:2026gge}. To the best of our knowledge, direct training of a quantum model on quantum hardware has not previously been demonstrated for jet classification.

In this paper, we explore the application of QGNNs to the problem of jet classification by constructing parametrized quantum circuits (PQCs) that respect the permutation symmetry of the jet's constituent particles. The model is trained on point cloud datasets for two classification tasks: quark versus gluon and up versus down jet discrimination, motivated by applications to the current LHC and future EIC collider experiments. We note that the up versus down jet discrimination task is expected to be more challenging than quark-gluon discrimination. The two flavors carry the same color representation, resulting in similar perturbative radiation patterns and making discrimination based on kinematic substructure alone particularly challenging~\cite{Larkoski:2019nwj}. As a result, this warrants the use of particle-level information such as particle identification (PID) and charge in addition to kinematic features. For the flavor tagging task, we incorporate these features, investigate their importance to the discriminative power of the QGNNs, and benchmark their performance against common jet observables and the PFN. Beyond ideal noiseless simulations, we also study the performance of a scaled-down model trained on Quantum Processing Units (QPUs). In summary, the main contributions of this work are twofold:

\begin{enumerate}
    \item A permutation-invariant QGNN for jet tagging, applied to quark versus gluon discrimination and, for the first time with a quantum model, light-quark ($u$ versus $d$) flavor tagging, benchmarked against the PFN and standard observables for each task;
    \item A demonstration of training and inference with the model on QPUs. We consider two architecturally distinct quantum computing platforms: the superconducting IBM Heron r2 and trapped-ion IonQ Forte-1 QPUs.
\end{enumerate}

The remainder of the paper is organized as follows. Section \ref{sec:qgnn_intro} defines and discusses the design of the QGNN. Section \ref{sec:setup} details the datasets used in the classification studies as well as the specific PQC designs used in each model. Section \ref{sec:results_ideal} presents the results of our idealized studies for the quark vs gluon and $u$ vs $d$ discrimination tasks. Section \ref{sec:hardware} presents the results of reduced QGNN models with noise simulators and on hardware backends. In the context of QML, Section \ref{sec:int} presents a new interpretability analysis that probes the correlations learned by the ideal quantum model. Finally, in Section \ref{sec:conclusion} we draw conclusions from these studies and present our outlook for further investigation.

\section{Quantum Graph Neural Networks} \label{sec:qgnn_intro}

\begin{figure}[b]
    \centering
    \includegraphics[width=0.5\linewidth]{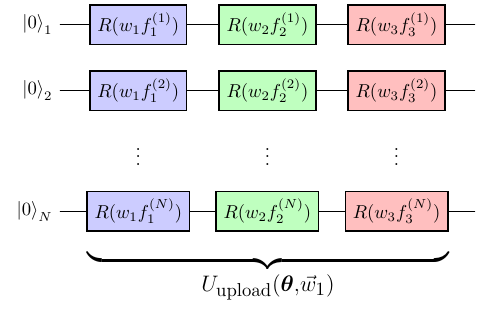}
    \caption{A single $N$-qubit encoding layer in which qubit $i$ receives
            angle-encoded features $\{f_1^{(i)}, f_2^{(i)}, f_3^{(i)}\}$ weighted by a shared uniform
            vector $\vec{w} = (w_1, w_2, w_3)$, constituting one upload block
            $U_{\mathrm{upload}}(\boldsymbol{\theta},\vec{w})$. Each qubit starts in the initial state $\ket{0}$.}
    \label{circ:encoding_full}
\end{figure}

In the Variational Quantum Classifier (VQC) approach to QML, a paradigm well suited to near-term quantum hardware~\cite{Cerezo:2020jpv}, a circuit consists of a sequence of parametrized quantum operations that process classical input data and produce a prediction through quantum measurement. In our approach, the classical input features are repeatedly encoded throughout the circuit using a data re-uploading scheme, with trainable quantum operations interleaved between successive data-encoding layers. This allows the circuit to progressively combine information from the input with learned transformations, increasing its expressive power compared with a single data-encoding stage. Measurements of the output quantum state are then used to obtain a prediction score.

Since the constituents of a jet carry no intrinsic ordering, this model should not depend on the arbitrary order in which those constituents are listed as input. In particular, let $\mathbf{x} = (x_1, \ldots, x_N)$ denote the per-particle feature vectors for the $N$ constituents fed into the model. For a permutation $\pi \in S_N$, where $S_N$ is the symmetric group acting on the $N$ jet constituents, define $\pi \cdot \mathbf{x} = (x_{\pi^{-1}(1)}, \ldots, x_{\pi^{-1}(N)})$ as the correspondingly reordered input. We require the full model $f_{\boldsymbol{\theta}}$, including the data encoding, the trainable quantum circuit, and the final measurement, to satisfy permutation invariance:
\begin{equation}
    f_{\boldsymbol{\theta}}(\mathbf{x}) = f_{\boldsymbol{\theta}}(\pi \cdot \mathbf{x}), \qquad \forall\, \pi \in S_N.
    \label{eq:model_perm_invariance}
\end{equation}

This symmetry is embedded by design in our model. We map the classical, particle-level information to qubits using a one-qubit-per-particle scheme, motivated by approaches in classical ML such as deep sets -- permutation-invariant neural networks designed to approximate functions on sets~\cite{Zaheer:2017wmg, Komiske:2018cqr}. Such particle-cloud-based approaches are found to outperform jet-image representations on jet classification tasks~\cite{Qu:2019gqs}. Under this scheme, permuting the particles simply permutes which qubit each particle's features are encoded onto, so the encoding map is equivariant under $S_N$ by construction. The features of the particles are angle-encoded, that is, encoded as rotations generated by Pauli operators ($R_x, R_y, R_z$). Some features are more expressive and discriminatory than others for specific classification problems, but it is sometimes difficult to know \textit{a priori} which features this may be. We therefore include a trainable weight for each encoded feature so the model may weight such features accordingly, improving the expressivity of the data encoding while giving the model the flexibility to learn scale as well as feature importance during optimization~\cite{Wach:2023ufx, Singh:2024dyh}. Crucially, these weights are shared across particles -- differing only across features and circuit layers -- so that they cannot be used to distinguish one particle from another, preserving equivariance.

In practice, encoding the classical data only once can restrict the class of functions representable by the circuit and therefore its expressibility~\cite{Schuld:2020enb, Perez-Salinas:2019pjx}. To alleviate this, we leverage the data reuploading technique, where each particle's features are periodically re-encoded but weighted by a different trainable vector $\vec{w}$. Here, ``periodically" implies that the reuploading occurs after each application of the ansatz, discussed next. The embedding scheme is shown in Figure \ref{circ:encoding_full}.

\begin{figure}[b]
    \centering
    \includegraphics[width=0.5\linewidth]{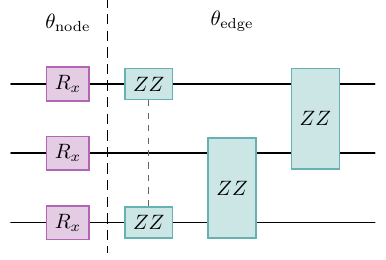}
    \caption{One layer $\mathcal{L}_{\bm{\theta}_k}(\bm{A})$ as defined in the main text for $N=3$ qubits. In the depicted layer, the node gates ($R_{x}$) are parametrized by $\theta_{\mathrm{node}}$ and the edge gates ($ZZ$) are parametrized by $\theta_{\mathrm{edge}}$. }
    \label{fig:ansatz_single_layer}
\end{figure}

The choice of ansatz is an important part of the circuit design. In classical machine learning, the choice of architecture determines the class of functions $\mathcal{F}$ that can be represented by the model. Similarly, in QML, the choice of ansatz, which includes the number and connectivity of qubits, the choice and arrangement of quantum gates, and the number of trainable parameters, plays an important role~\cite{Schuld:2020enb}. The choice of ansatz also embeds inductive biases that can considerably simplify the learning process and allow the model to generalize better. In the context of jet classification, an important inductive bias is permutation symmetry: here, we choose the ansatz to be \emph{equivariant} under qubit relabeling for the model to satisfy the invariance condition of Eq.~\eqref{eq:model_perm_invariance}, so a graph-like ansatz, where the particles are represented by nodes, is well-suited. Following similar constructions to Refs.~\cite{Schatzki:2022tfq, Mernyei:2021krm}, we define our quantum graph ansatz as follows:

\newtheorem*{eqgansatz}{Quantum Graph Ansatz:}

\begin{eqgansatz}\label{def:gate_ansatz}
Given an adjacency matrix $\boldsymbol{A}$, it is a parameterized family of quantum circuits
$C_{\boldsymbol{\theta}}(\boldsymbol{A})$ acting on $N$ qubits, defined as a composition of $L$ layers:
\[
C_{\boldsymbol{\theta}}(\boldsymbol{A})
=
\mathcal{L}_{\boldsymbol{\theta}_L}(\boldsymbol{A})
\cdots
\mathcal{L}_{\boldsymbol{\theta}_1}(\boldsymbol{A}),
\]
where each layer is a graph-conditioned unitary of the form
\[
\mathcal{L}_{\boldsymbol{\theta}_k}(\boldsymbol{A})
=
\left(
\prod_{i=1}^{N} U^{(\mathrm{node})}_{k}(\theta_{\mathrm{node},k})
\right)
\left(
\prod_{(l,m)\in E(\boldsymbol{A})} U^{(\mathrm{edge})}_{k}(\theta_{\mathrm{edge},k})
\right),
\]
where $E(\boldsymbol{A}) = \{(l,m) \mid A_{lm}=1, l<m\}$ denotes the set of undirected edges, with $l,m\in\{1,\ldots,N\}$ labeling the graph nodes (qubits). Here, $U^{(\mathrm{node})}_{k}(\theta_{\mathrm{node},k})$ is the same single-qubit gate applied to every node at layer $k$, and $U^{(\mathrm{edge})}_{k}(\theta_{\mathrm{edge},k})$ is the same two-qubit entangling gate applied to every edge at layer $k$, subject to the symmetry and commutativity conditions detailed below. The layer parameters are $\boldsymbol{\theta}_k = (\theta_{\mathrm{node},k}, \theta_{\mathrm{edge},k})$.
\end{eqgansatz}

We require that $U^{(\mathrm{edge})}$ be symmetric under exchange of the two qubits it acts on and must commute across all edges, including edges that share a qubit. Figure \ref{fig:ansatz_single_layer} is an example of a single layer of the ansatz for such a graph. For the purposes of our study, we examine fully-connected graphs, so $A_{ij} = 1- \delta_{ij}$ defines $\boldsymbol{A}$. Note that under a permutation $\hat{P}$ of qubits (or nodes in the language of graphs), $\boldsymbol{A} \rightarrow \hat{P}\boldsymbol{A}\hat{P}^T$, i.e. the adjacency matrix transforms similarly such that permutation symmetry is not broken. The layer of single qubit gates is also evidently invariant under this transformation. Therefore, under the encoding scheme introduced earlier, the quantum circuit designed thus far is equivariant under $S_N$, as required by Eq.~\eqref{eq:model_perm_invariance}; the remaining ingredient, an $S_N$-invariant readout, is introduced in the aggregation step described next.

\begin{figure}[t]
    \centering
    \includegraphics[width=\linewidth]{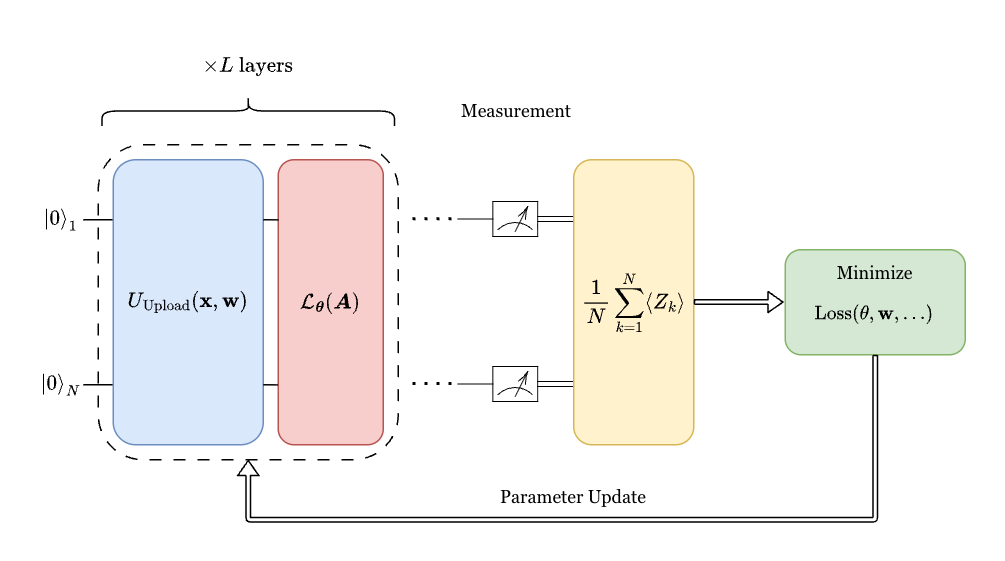}
    \caption{A summary of our Quantum Graph Neural Network.}
    \label{fig:full_diagram}
\end{figure}

Finally, we define the aggregation function in the $N$-qubit quantum graph model as follows:
\[
g = \frac{1}{N} \sum_{i=1}^{N} \langle Z_i \rangle\,,
\]
where \(\langle Z_i\rangle\) denotes the expectation value of the Pauli-\(Z\) operator measured on qubit \(i\). This quantity provides a scalar summary of the quantum state that is invariant under permutations of the qubit indices. This choice is motivated by classical graph neural networks, where node features are aggregated using symmetric functions, ensuring permutation invariance with respect to the ordering of the input features. Common choices include $\mathrm{SUM}$, $\mathrm{MEAN}$, and $\mathrm{MAX}$, with the element-wise mean being a particularly widespread choice, for example in Dynamic Graph Convolutional Networks such as ParticleNet~\cite{Qu:2019gqs}. We summarize our architecture in Figure \ref{fig:full_diagram}.  

\section{Methodology}\label{sec:setup}

\subsection{Event Generation}

We utilize two datasets for the two studies explored in the paper. Since our results are intended to be relevant to the LHC and the EIC, each dataset is generated under the respective collision conditions.  

First, for the quark-gluon study, we use the quark and gluon jet dataset provided by the \texttt{EnergyFlow} library~\cite{komiskePythia8QuarkGluon2019,Komiske:2018cqr}. The samples are generated from the processes
$q\bar{q} \to Z(\to \nu\bar{\nu}) + g$
and $qg \to Z(\to \nu\bar{\nu}) + (u,d,s)$
in $pp$ collisions at $\sqrt{s}=14~\mathrm{TeV}$. Hadronization and multiple parton interactions are enabled using the default tunings and shower parameters. Final-state non-neutrino particles are clustered into anti-$k_T$ jets with radius parameter $R=0.4$ using \textsc{FastJet}~3.3.0~\cite{Cacciari:2011ma}. Jets are required to satisfy transverse momentum $p_T \in [500,550]~\mathrm{GeV}$ and rapidity $|\eta|<1.7$. No detector simulation is performed. From these jets, we only retain jets initiated by up, down, or strange quarks for the quark class. Up to 150 constituent particles sorted in decreasing $p_T$ are included in the event description. 

We retain the $(z, \eta, \phi)$ information for each particle. To ensure the models are more sensitive to momentum distribution than energy scale, the fractional transverse momentum $z_i = p_{T,i}/p_{T,\mathrm{Jet}}$ is used instead of raw per-particle $p_T$. Additionally, the $\phi$ values are rescaled to values in $[0,1]$. A total of 5000 jets are used, with an 80-20 training/testing split. 

Second, for the up-down flavor study, we use the flavor-tagging dataset introduced in Refs.~\cite{Lee:2022kdn, leePYTHIA6DatasetMachine2023}. The jet samples are generated using leading-order (LO) deep inelastic scattering (DIS) as the hard-scattering process, where the final state consists of the scattered electron and a single jet originating from different quark flavors. The underlying LO DIS process is $\gamma^* q \to q$, implemented in PYTHIA6 as process 99. Jet flavor is identified using the flavor of the underlying quark in the LO DIS process. Events are required to satisfy photon virtuality and inelasticity ranges of $25 < Q^2 < 1000~\mathrm{GeV}^2$ and $0.1 < y < 0.85$, respectively. 

We retain the $(z, \eta, \phi, \mathrm{PID}, q)$ information for each particle in the jet, where $q$ is the charge of the particle and PID is the particle's PDG ID~\cite{ParticleDataGroup:2024cfk}. The unique PDG IDs are sorted and mapped to values spaced by 0.1 and centered around 0. For example, $-11$ (positron) is mapped to $-0.1$ and $11$ (electron) is mapped to $0.1$. The $\phi$ values are also rescaled to values in $[0,1]$. A total of 5000 jets are used in both training and evaluation of the quantum models with an 80-20 split.

\subsection{Training and Evaluation}

Depending on the event conditions and clustering algorithm, a jet may contain an arbitrarily large number of particles. However, for the data encoding scheme we consider in Section \ref{sec:qgnn_intro}, not all particles in the jet can be included in the QGNN since a jet may contain too many particles to assign qubits individually. A suitable truncation scheme is therefore required for a computationally tractable model. We constrain the number of particles used in the jet to the $N$-hardest (i.e. highest-$p_T$) particles in the jet, where $N$ can be varied. This choice retains the dominant momentum-carrying constituents of the jet which capture much of its energetic structure and can preserve substantial jet-classification information despite retaining only a subset of constituents~\cite{Vigl:2026ppx, Dasgupta:2013ihk}.

\begin{figure}[t]
    \centering

    \begin{subfigure}[b]{0.49\textwidth}
        \centering
        \includegraphics[width=\textwidth]{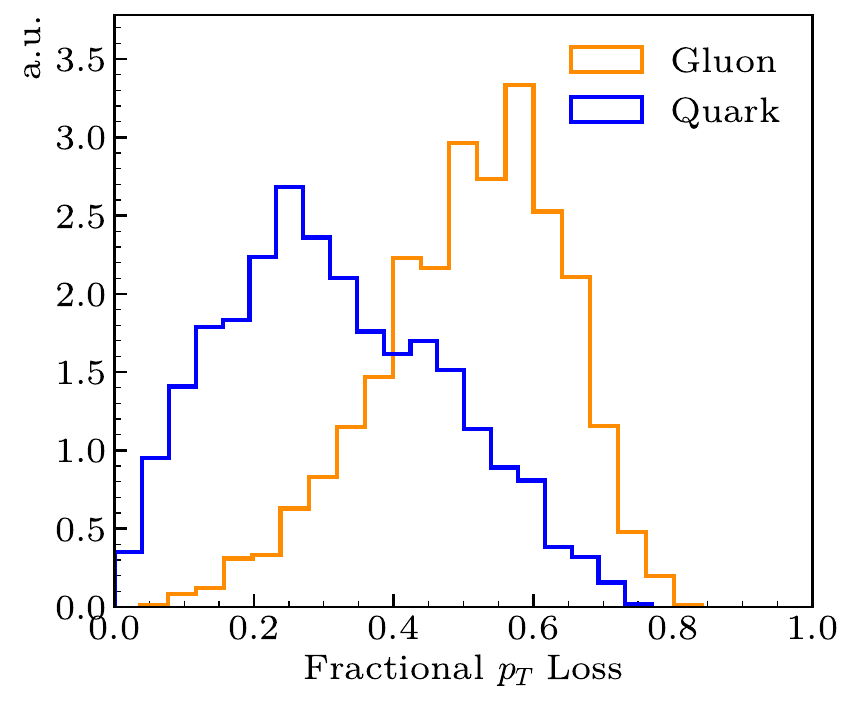}
        \label{fig:sub1_0}
    \end{subfigure}
    \begin{subfigure}[b]{0.49\textwidth}
        \centering
        \includegraphics[width=\textwidth]{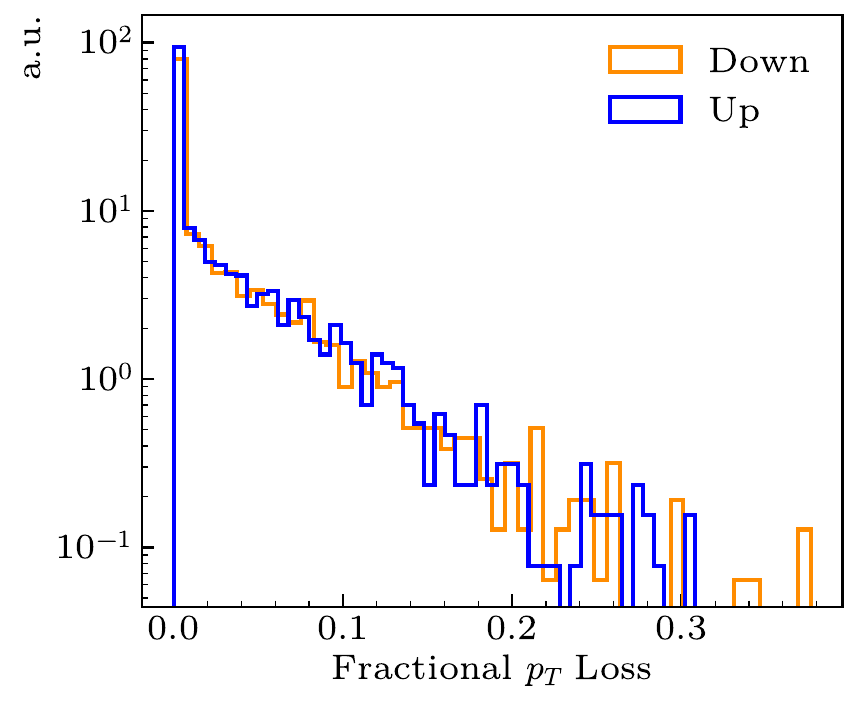}
        \label{fig:sub2_0}
    \end{subfigure}

    \caption{Total removed fraction of $p_T$ for each jet as a result of the 5-particle and 10-particle truncation schemes for the quark-gluon and flavor models respectively. A larger $p_T$ loss indicates that more of the jet's transverse momentum is lost post-truncation.}
    \label{fig:removed_pt}
\end{figure}

We begin by detailing the quark-gluon study. The dataset is truncated to the hardest $5$ particles in the jet. Since the quark-gluon dataset contains three features, $(z, \eta, \phi)$, per particle, each data encoding layer $U_{\mathrm{upload}}(\mathbf{\theta},\vec{w})$ is weighted by a 3-element vector $\vec{w}$. We use $L=6$ parametrized $\mathcal{L}_{\mathbf{\theta}}$ ansatz layers. Therefore, the total number of trainable parameters in the model is $3L + 2L = 30$. Here, 3 is the contribution from the data encoding layer which applies a trainable weight for each feature, and 2 is the contribution from the graph ansatz for single and two qubit parametrized gates. 

For the flavor study, the dataset is truncated to the hardest 10 particles in the jet. Since the dataset contains five features, $(z, \eta, \phi, \mathrm{PID}, q)$, per particle, each upload layer $U_{\mathrm{upload}}(\mathbf{\theta},\vec{w})$ is weighted by a 5-element vector $\vec{w}$. We use $L=6$ parametrized $\mathcal{L}_{\mathbf{\theta}}$ ansatz layers. Therefore, the total number of trainable parameters in the model is $5L + 2L = 42$, where 5 and 2 are contributions from the data encoding and ansatz layers respectively. The effect of the jet truncation for each dataset is shown in Figure \ref{fig:removed_pt}. 
Because quark jets generally have lower constituent multiplicities, the fraction of jet transverse momentum $p_T$ lost through truncation is smaller for quark jets than for gluon jets, which typically have higher multiplicities and softer fragmentation~\cite{Gallicchio:2011xq,Larkoski:2019nwj}.

Both the quark–gluon and flavor models are trained using $50$ epochs with a mini-batch size of $50$. The model is trained using the hinge loss applied to the expectation values of the quantum circuit. Given the model output for a training jet $j$, $g(j) \in [-1, 1] $ and a binary truth label $y_j$, the resulting loss is
\[
{L}_{\mathrm{hinge}} = \frac{1}{N_\mathrm{Jets}} \sum_{j=1}^{N_\mathrm{Jets}} \max\!\left(0,\, 1 - \tilde{y}_j g_j \right), 
\quad
\tilde{y}_j = 2y_j - 1.
\]

\noindent Optimization is performed using the \textsc{Adam} optimizer~\cite{Kingma:2014vow} implemented via \texttt{optax}, built in JAX~\cite{deepmind2020jax}, which provides automatic differentiation, just-in-time (JIT) compilation, and vectorized evaluation of the quantum circuits. The learning rate is set to $0.01$ for both the quark–gluon model and the flavor model.

We evaluate each model using a receiver-operating characteristic (ROC) curve. For a classifier producing scores on each sample it is evaluated on, the ROC visualizes the true positive rate (TPR) and false positive rate (FPR) at various thresholds by plotting the TPR as a function of the FPR:

\[
\mathrm{TPR} = \frac{\mathrm{TP}}{\mathrm{TP}+\mathrm{FN}}, \qquad \mathrm{FPR} = \frac{\mathrm{FP}}{\mathrm{FP}+\mathrm{TN}},
\]
where TP, TN, FP, and FN denote the numbers of true positives, true negatives, false positives, and false negatives respectively at a given threshold. Equivalently, the TPR and FPR are the signal ($\varepsilon_S$) and background efficiencies ($\varepsilon_B$) respectively. By varying the classification threshold, the ROC curve therefore characterizes the trade-off between signal efficiency and background rejection for the classifier. The area under the ROC curve (AUC) provides a threshold-independent summary of the classifier performance. An AUC value of 1 corresponds to perfect discrimination between the two jet classes, whereas an AUC value of 0.5 indicates performance equivalent to random chance. Further, we compute uncertainty estimates for the model by retraining with 10 different initial seeds for both the model and dataset. A 95\% confidence interval is constructed pointwise along the ROC curve, in addition to a confidence interval for the average AUC.

We additionally evaluate classifier performance using the significance improvement characteristic (SIC)~\cite{Gallicchio:2012ez}, defined as
\[
\mathrm{SIC} = \frac{\varepsilon_S}{\sqrt{\varepsilon_B}},
\]
where $\varepsilon_S$ and $\varepsilon_B$ denote the signal and background efficiencies, respectively. The SIC quantifies the improvement in the statistical significance of a signal over background after applying a selection cut, and is therefore useful for identifying thresholds for the model that maximize signal sensitivity. We note that both the ROC and SIC metrics are independent of absolute event yields or integrated luminosities, allowing straightforward comparison across different models.

\subsection{Performance Benchmarks}

Classical jet tagging approaches include the use of hand-crafted observables or Deep Neural Networks acting on jet representations such as particle clouds. To characterize the performance of the quantum model, we compare its performance to such classical approaches, namely the PFN architecture~\cite{Komiske:2018cqr} and observables robust to each classification task. The PFN is a DeepSets architecture~\cite{Zaheer:2017wmg} that can operate on variable-length jet descriptions and is permutation invariant like our proposed quantum model. Conceptually, the permutation-invariance of the PFN is expressed as $f(p_1, \ldots, p_M) = f(p_{\pi(1)}, \ldots, p_{\pi(M)})$, where $\pi \in S_M$, and 

\[
f(p_1, \ldots, p_M) = F\left( \sum_{i=1}^M \Phi(p_i)\right).
\]

The PFN takes as input an $n$-dimensional particle-level feature vector $p$, where $p$ can include the kinematic features of the particle as well as ID. $\Phi : \mathbb{R}^n \to \mathbb{R}^d$ here represents a per-particle network that maps each particle to a $d$-dimensional latent space. $F: \mathbb{R}^d \to \mathbb{R}$ is a network that maps the $d$-dimensional representation to the model output. This construction of $f$ makes permutation invariance manifest. 

For each classification task, we tune the hyper-parameters of the PFN. In particular, we adjust the depth of $\Phi, F$, the number of neurons in each layer, and training parameters such as epochs and batch size to achieve the best performance on the associated dataset. The PFN is trained with the Adam optimizer and binary cross-entropy loss. We apply a ReLU activation to each neuron of the dense layers of the network and a softmax output activation function. The PFN is trained for 50 epochs and a batch size of 200 samples.  

For the quark-gluon study, we consider the jet girth observable~\cite{Almeida:2008yp, Yan:2020zrz} defined below:

\[
\mathrm{\mathbf{Jet \ Girth:}} \quad g_J =  \sum_{i\in J} \frac{p_{T, i}}{p_{T, J}} \Delta R_{i},
\]
where $\Delta R_{i} = \sqrt{\Delta \eta_i^2 + \Delta \phi_i^2}$ is the distance of the particle in $(\eta,\phi)$ space from the jet axis. We note that jet girth belongs to the wider class of jet angularities ($\lambda^\kappa_\beta$)~\cite{Reichelt:2021svh} which probe the angular and transverse momentum distributions of the particles within a jet. In particular, a lower value of jet girth for a given $p_{T, J}$ corresponds to a more collimated jet. In the soft collinear limit, emission probabilities scale as $\alpha_s C_R$, where $C_R = C_F = 4/3$ for quark-initiated jets and $C_R = C_A = 3$ for gluon-initiated jets. A larger overall color factor contributes to the higher average multiplicity of gluon jets compared to quark jets and, as a result, a wider radiation profile. Therefore, we expect that jet girth is a strong discriminator for this particular discrimination task.  

\begin{figure}[t]
    \centering
    \includegraphics[width=1.01\linewidth]{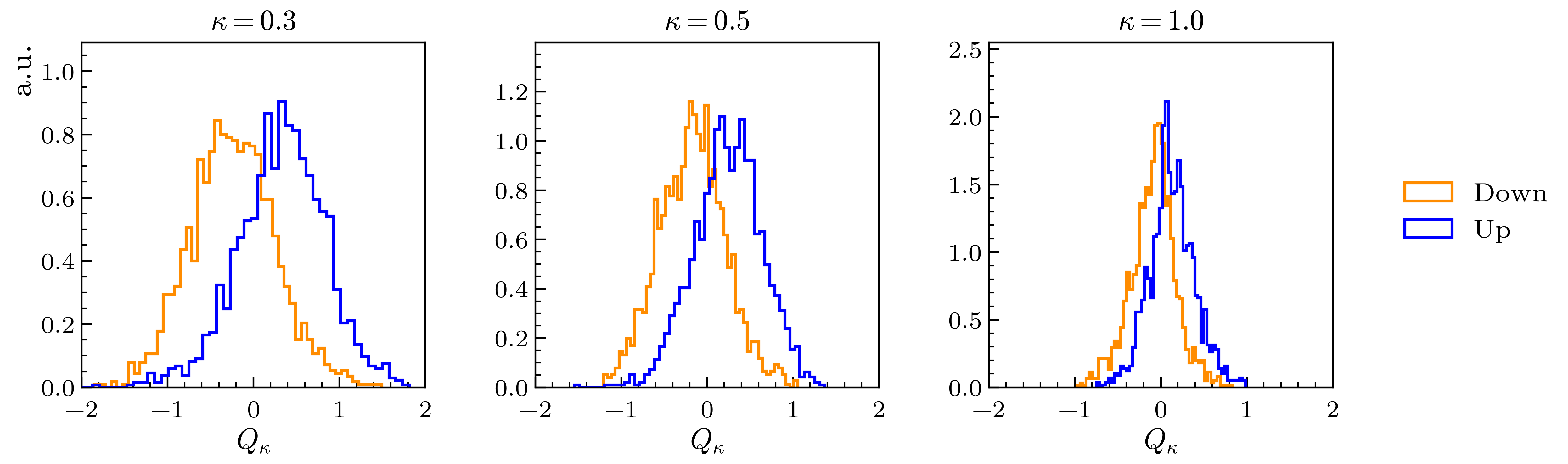}
    \caption{Jet Charge distributions for up (blue) and down (yellow) jets for $\kappa = 0.3$ (left), $\kappa = 0.5$ (middle), and $\kappa = 1.0$ (right).}
    \label{fig:jc_dist_allkappa}
\end{figure}

\begin{figure}[t]
    \centering

    \begin{subfigure}[b]{0.49\textwidth}
        \centering
        \includegraphics[width=1.125\textwidth]{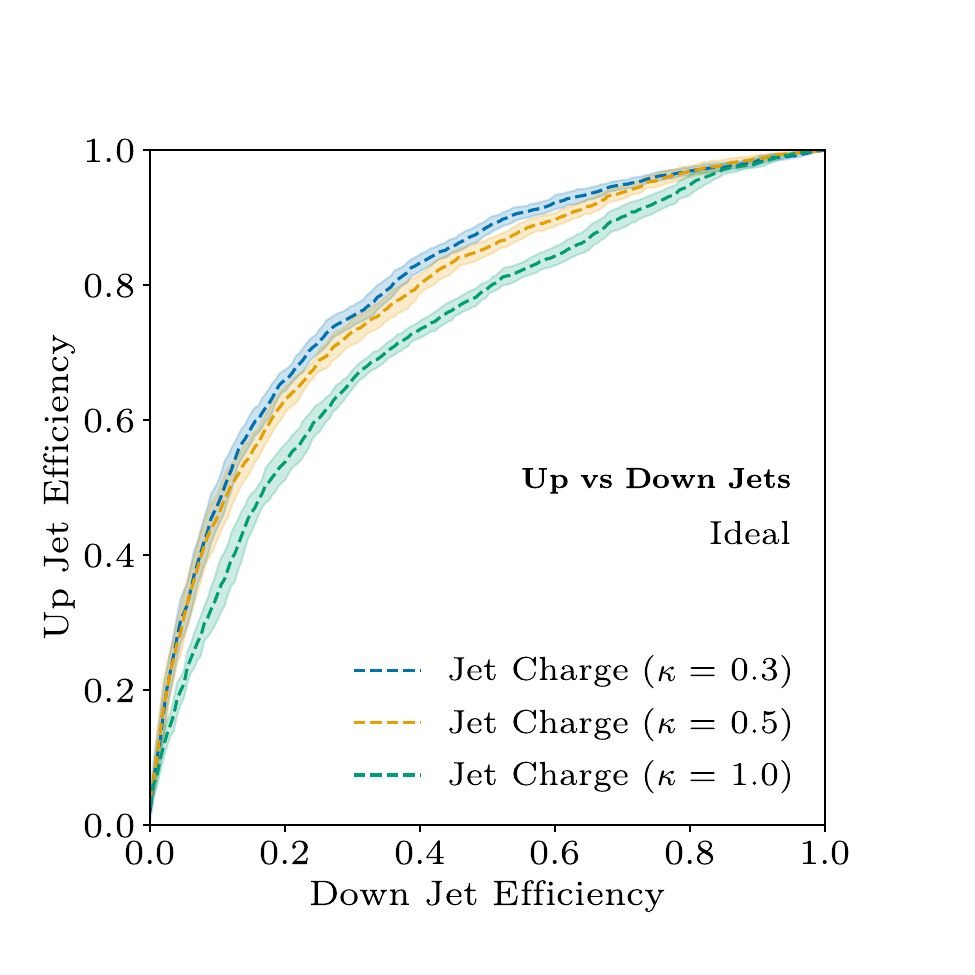}
        \label{fig:sub1_1}
    \end{subfigure}
    \begin{subfigure}[b]{0.49\textwidth}
        \centering
        \includegraphics[width=1.125\textwidth]{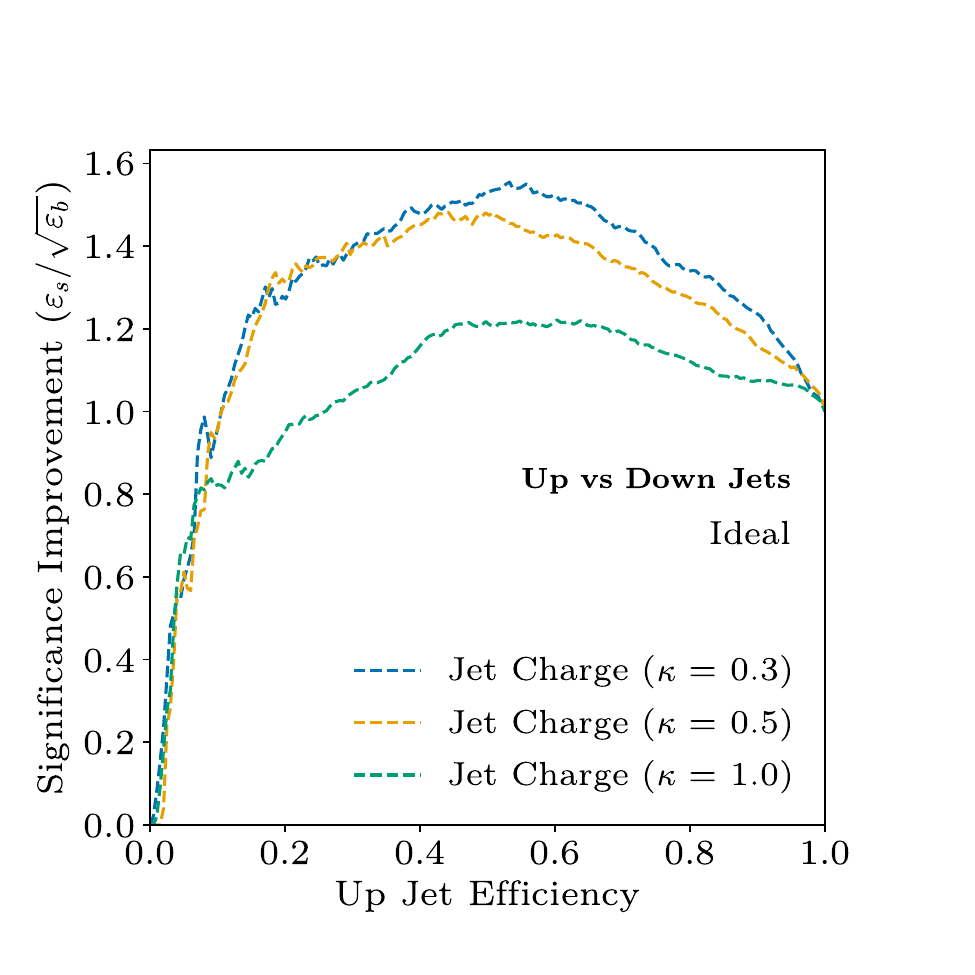}
        \label{fig:sub2_1}
    \end{subfigure}
    \caption{The ROC (left) and SI (right) curves for jet charge evaluated at different values of the momentum-weighting parameter $\kappa$. Smaller values of $\kappa$ provide stronger discrimination, with $\kappa=0.3$ yielding the best overall performance. The SI curves demonstrate that the choice of $\kappa$ affects the achievable significance improvement.}
    \label{fig:jcs_roc_sic}
\end{figure}

For the flavor study, we consider the jet charge observable~\cite{Field:1977fa, Lee:2022kdn, Krohn:2012fg, Kang:2023ptt} defined below:

\[
\mathrm{\mathbf{Jet \ Charge:}} \quad Q_{\kappa} =  \sum_{i\in J} \left(\frac{p_{T, i}}{p_{T, J}} \right)^{\kappa} q_i.
\]
Typically, values of $\kappa$ between 0.2 and 1 are used to weight the particle's transverse momentum fraction ($p_{T, i}/p_{T, J}$)~\cite{Berge:1980dx, ALEPH:1991fba}. Figure \ref{fig:jc_dist_allkappa} shows the distributions of the jet charge for both up and down jets. Additionally, we compare the performance of jet charge for each of $\kappa \in \{0.3, 0.5, 1\}$ in Figure \ref{fig:jcs_roc_sic}. These results demonstrate that jet charge is indeed a strong discriminator for up and down light quark jets. Generally, one finds good separation for $\kappa \sim 0.3$. Early applications of jet charge focused on distinguishing quark jets and probing the electroweak properties of quarks in deep-inelastic scattering and $e^+e^-$ collisions~\cite{Berge:1980dx}. It has more recently gathered interest for tagging at the LHC~\cite{ATLAS:2015rlw,Li:2019dre}, for example, in quark and gluon jet separation. Its flavor sensitivity is especially valuable at the EIC, where jet charge can serve as a ``flavor prism'' for spin asymmetries and help constrain the flavor dependence of nucleon structure~\cite{Kang:2020fka}. Extensions such as dynamic jet charge provide complementary flavor-tagging and nuclear-structure sensitivity~\cite{Kang:2021ryr}, while jet charge in electron-nucleus collisions has also been proposed as a probe of neutron-skin thickness~\cite{Zhang:2025raf}.

\section{Results: Ideal Simulation}\label{sec:results_ideal}

\subsection{Quark vs Gluon Tagging}

\begin{figure}[t]
    \centering

    \begin{subfigure}[b]{0.49\textwidth}
        \centering
        \includegraphics[width=1.125\textwidth]{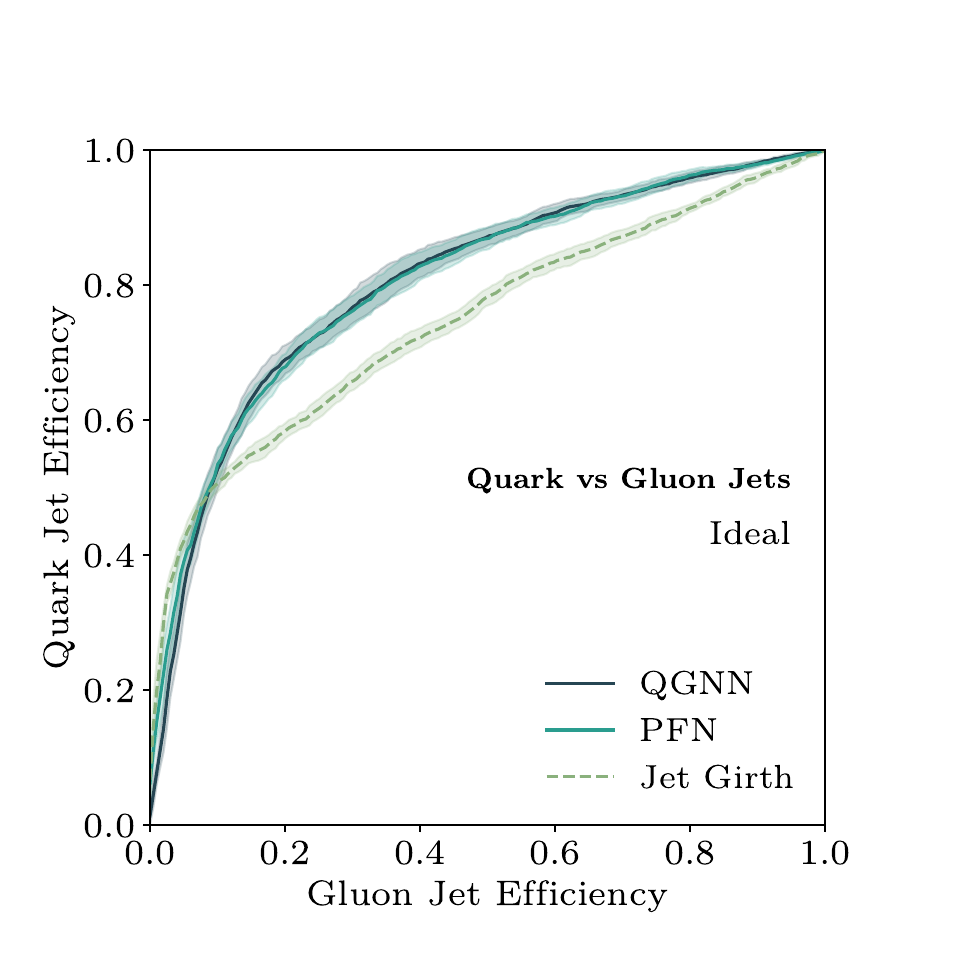}
        \label{fig:sub1_2}
    \end{subfigure}
    \begin{subfigure}[b]{0.49\textwidth}
        \centering
        \includegraphics[width=1.125\textwidth]{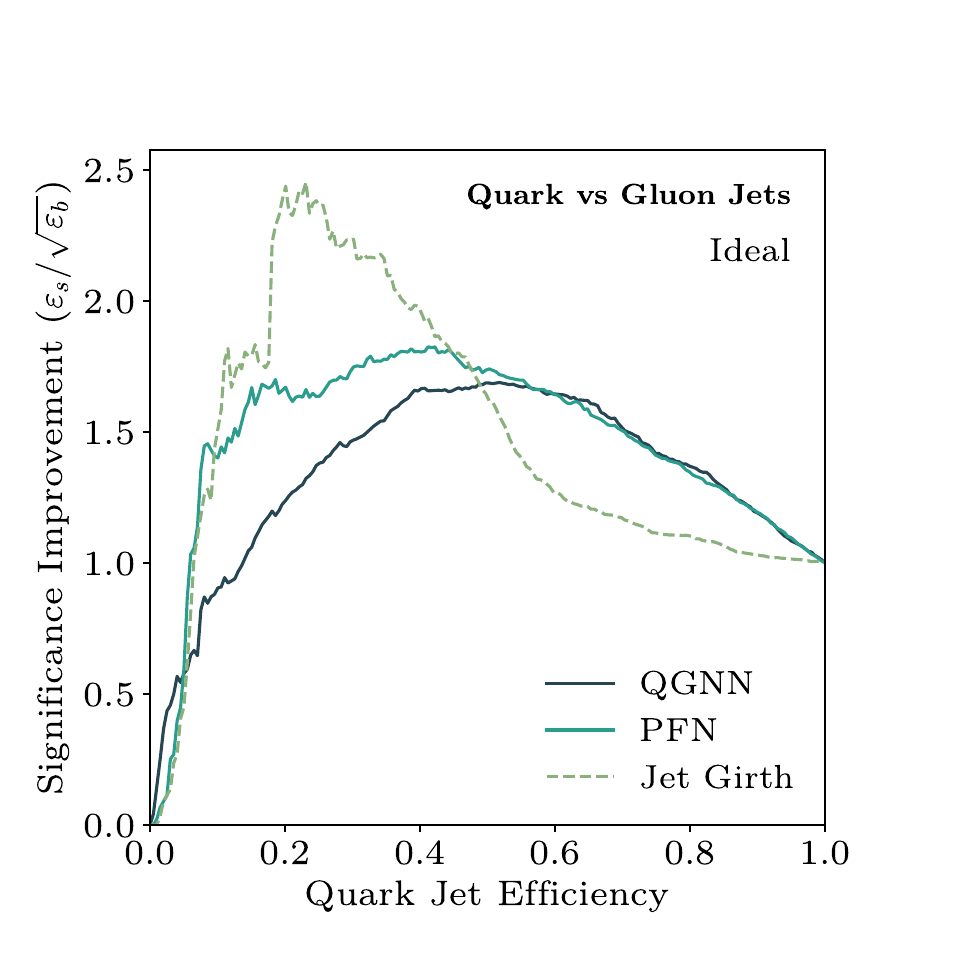}
        \label{fig:sub2_2}
    \end{subfigure}

    \caption{The ROC (left) and SI (right) curves for the QGNN-QG, PFN, and jet girth observable. The QGNN-QG and PFN models perform roughly equally and have better AUC than the jet girth observable.}
    \label{fig:qg_roc_sic}
\end{figure}

\begin{table}[b]
\centering
\begin{tabular}{|l|l|c|}
\hline
\textbf{Model} & \textbf{Description} & \textbf{AUC} \\
\hline
\noalign{\vskip 2pt}
\hline
\textbf{QGNN-QG} & QGNN on particle three-momenta & \textbf{0.806 $\pm$ 0.012} \\
PFN & PFN on particle three-momenta & 0.804 $\pm$ 0.012 \\
Jet Girth & Kinematic jet shape observable & 0.747 $\pm$ 0.007 \\
\hline
\end{tabular}
\caption{Quark vs gluon jet discrimination performance for different models/jet observables over 10 different seeds. The QGNN-QG model is found to perform competitively with each approach.}
\label{tab:qg_quc_scores}
\end{table}

The PFN is given $p_i = (z_i, \eta_i, \phi_i)$ for each particle in each jet and is trained and evaluated on the same dataset as the quantum model. We use two $\Phi$ layers with 150 and 400 neurons, and two $F$ layers with 300 neurons each. Similar to the quantum model, we apply the PFN to the truncated dataset consisting of the 5 hardest particles of each jet. 

\begin{figure}[t]
\centering
    \begin{subfigure}{0.49\linewidth}
        \centering
        \includegraphics[width=\linewidth]{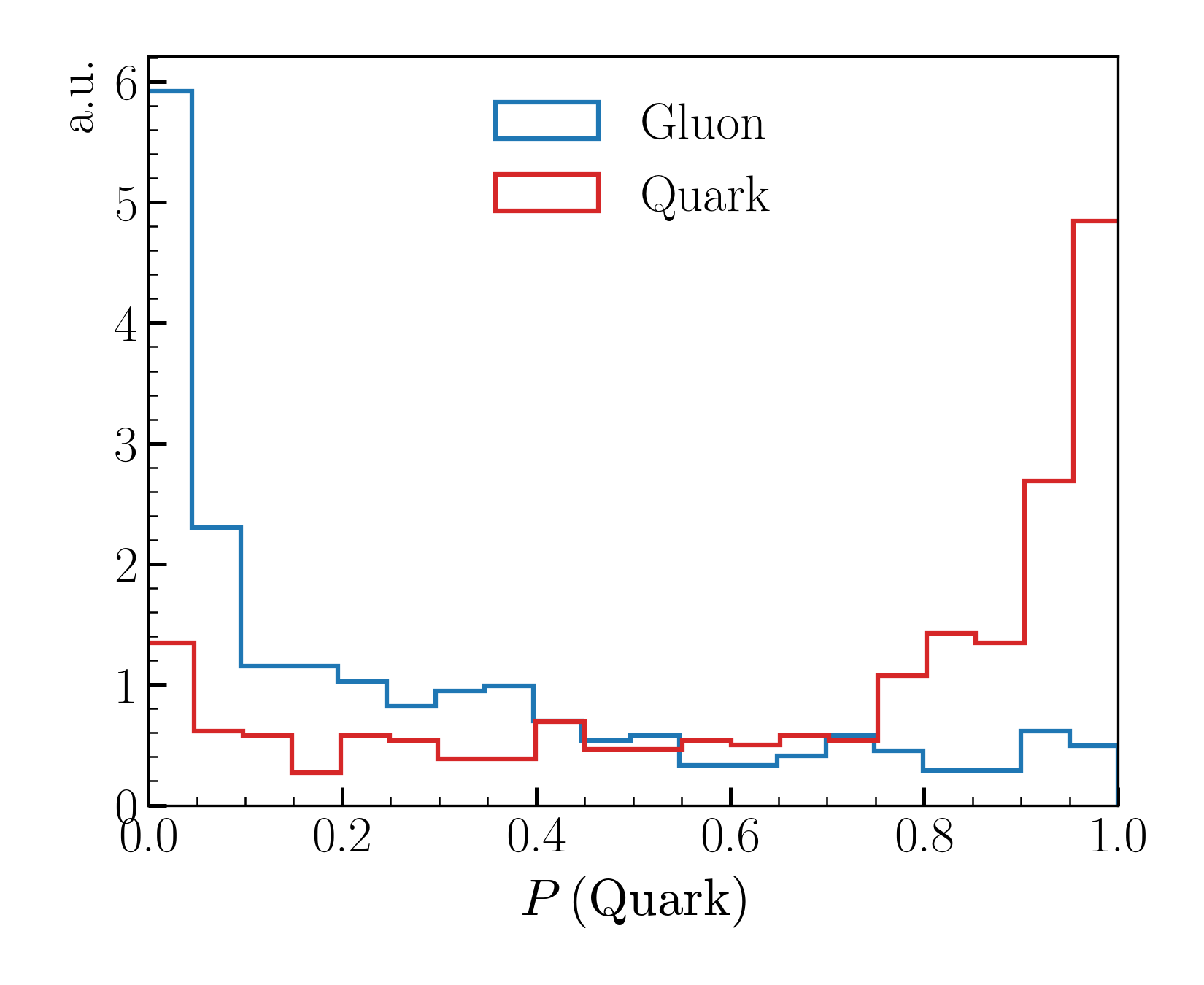}
    \end{subfigure}
    \hfill
    \begin{subfigure}{0.49\linewidth}
        \centering
        \includegraphics[width=\linewidth]{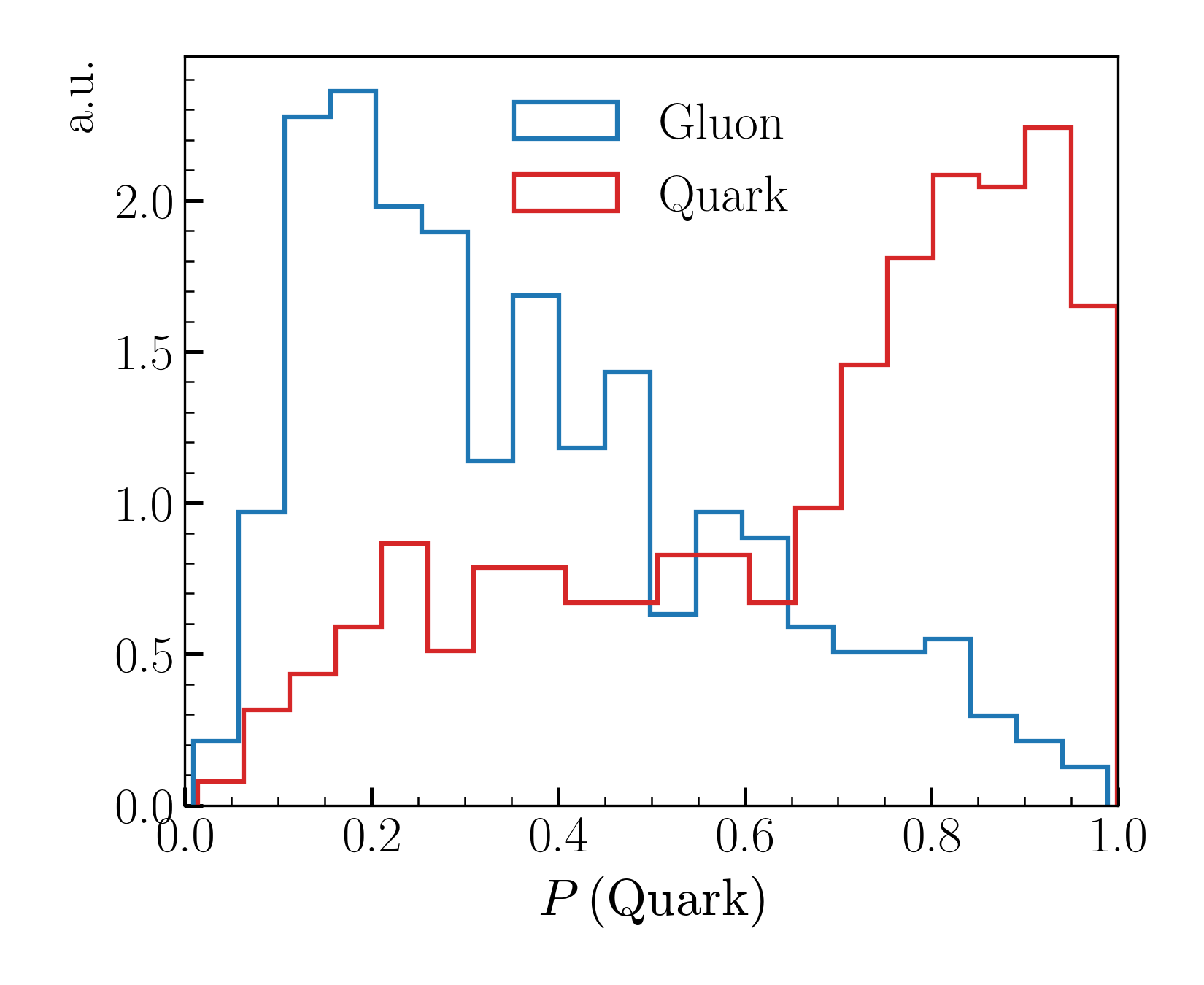}
    \end{subfigure}
    
    \caption{QGNN-QG and PFN discrimination histograms on evaluation datasets. (Left) QGNN discriminator output. (Right) PFN discriminator output.}
    \label{fig:qg_disc_plots}

\end{figure}

Figure \ref{fig:qg_roc_sic} depicts the performance of the QGNN-QG compared to the PFN model, and the corresponding AUC values are listed in Table \ref{tab:qg_quc_scores}. We find that the QGNN-QG is able to perform competitively compared to the classical PFN and jet girth observables. Specifically, the PFN and QGNN-QG perform comparably while $g_J$ is relatively worse. The SI curve demonstrates that a significance improvement of at least 1.5 can be achieved by the model. It is desirable for a model to have a higher significance improvement. Figure~\ref{fig:qg_disc_plots} compares the discriminative power of both QGNN-QG and the PFN for a given seed instance, where the models are found to separate both classes well.

\subsection{Flavor Tagging}

\begin{figure}[t]
    \centering

    \begin{subfigure}[b]{0.49\textwidth}
        \centering
        \includegraphics[width=1.125\textwidth]{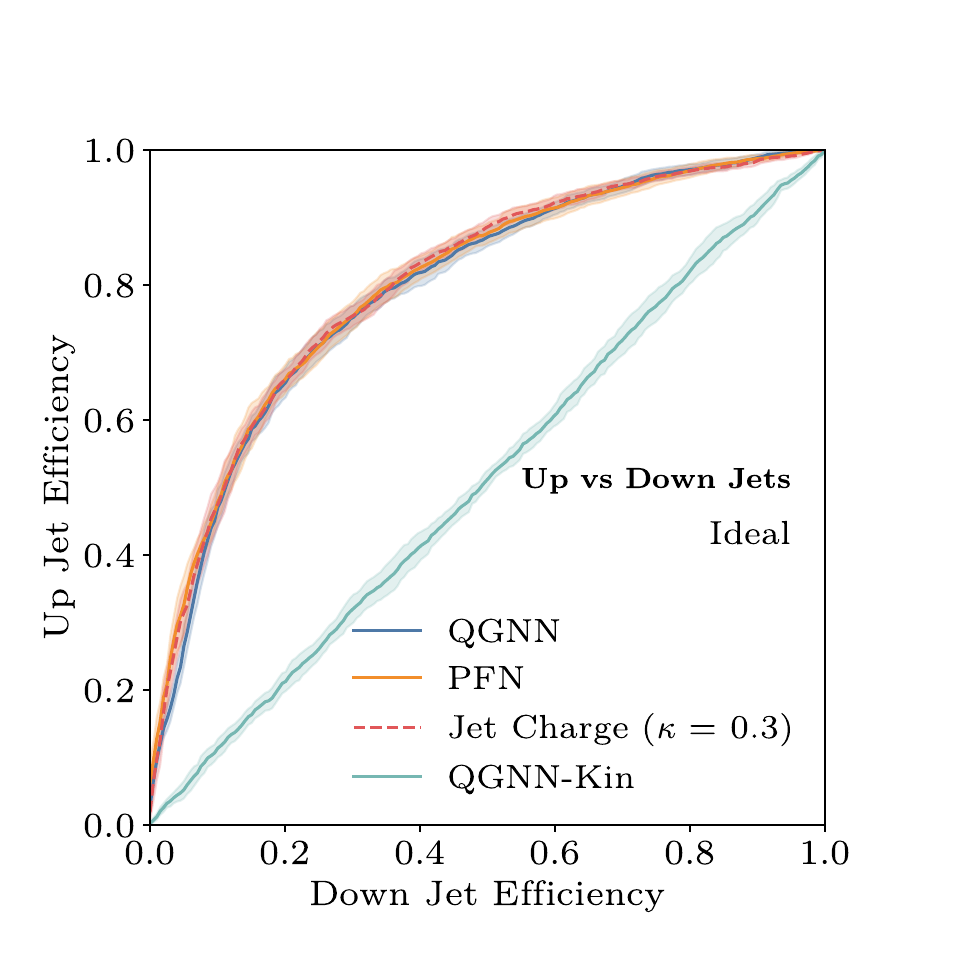}
        \label{fig:sub1_3}
    \end{subfigure}
    \begin{subfigure}[b]{0.49\textwidth}
        \centering
        \includegraphics[width=1.125\textwidth]{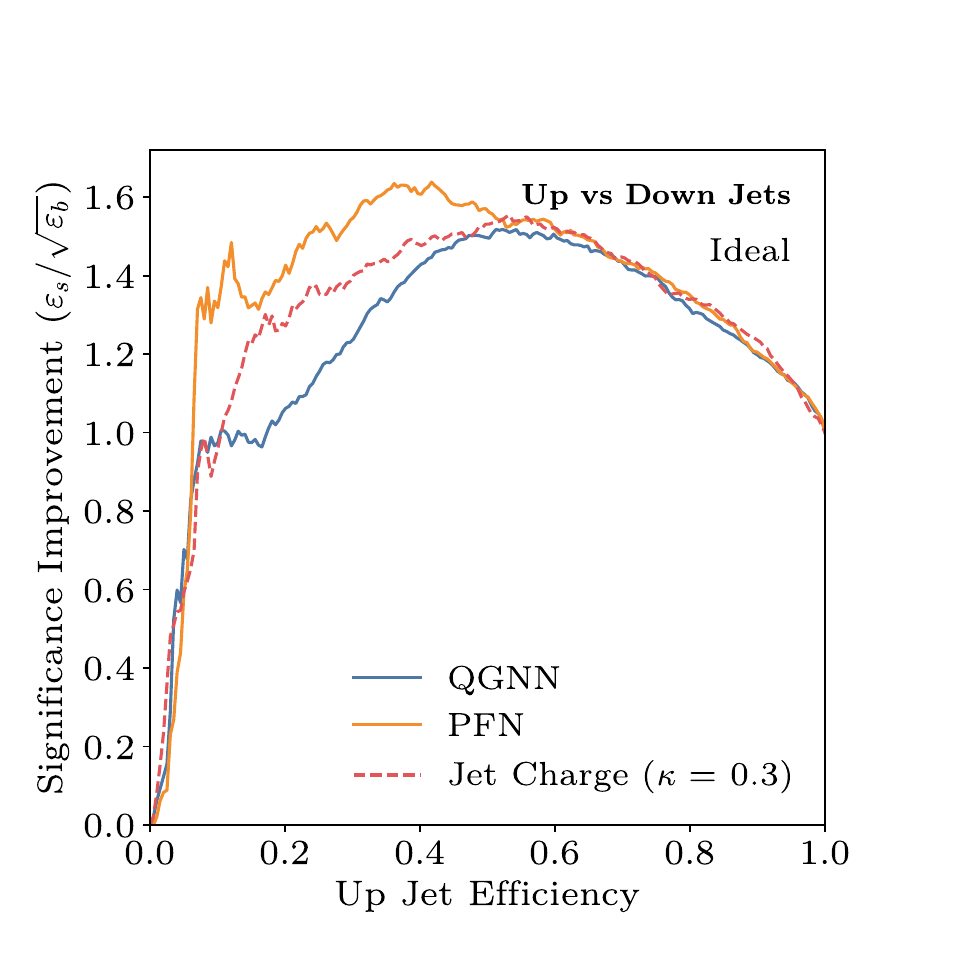}
        \label{fig:sub2_3}
    \end{subfigure}

    \caption{The ROC (left) and SI (right) curves for the QGNN-UD, PFN, $Q_{\kappa=0.3}$, and QGNN-Kin in the up-versus-down jet classification task. The first three approaches exhibit comparable discrimination power, with jet charge performing competitively with the learned classifiers. As expected, the QGNN-Kin classifier with purely kinematic information fails to distinguish between the two flavors. The SI curves indicate that the different methods achieve similar significance improvement across a broad range of signal efficiencies.}

    \label{fig:ud_roc_sic}
\end{figure}

Similarly to the quark-gluon study, the PFN model is provided $(z, \eta, \phi, \mathrm{PID}, Q)$ for each particle in each jet and trained and evaluated on the same dataset as the quantum model. We tune the PFN model with two $\Phi$ layers with 400 and 30 neurons and four $F$ layers with 10, 480, 480, and 20 neurons respectively.

\begin{figure}[t]
\centering

\begin{subfigure}{0.49\linewidth}
    \centering
    \includegraphics[width=\linewidth]{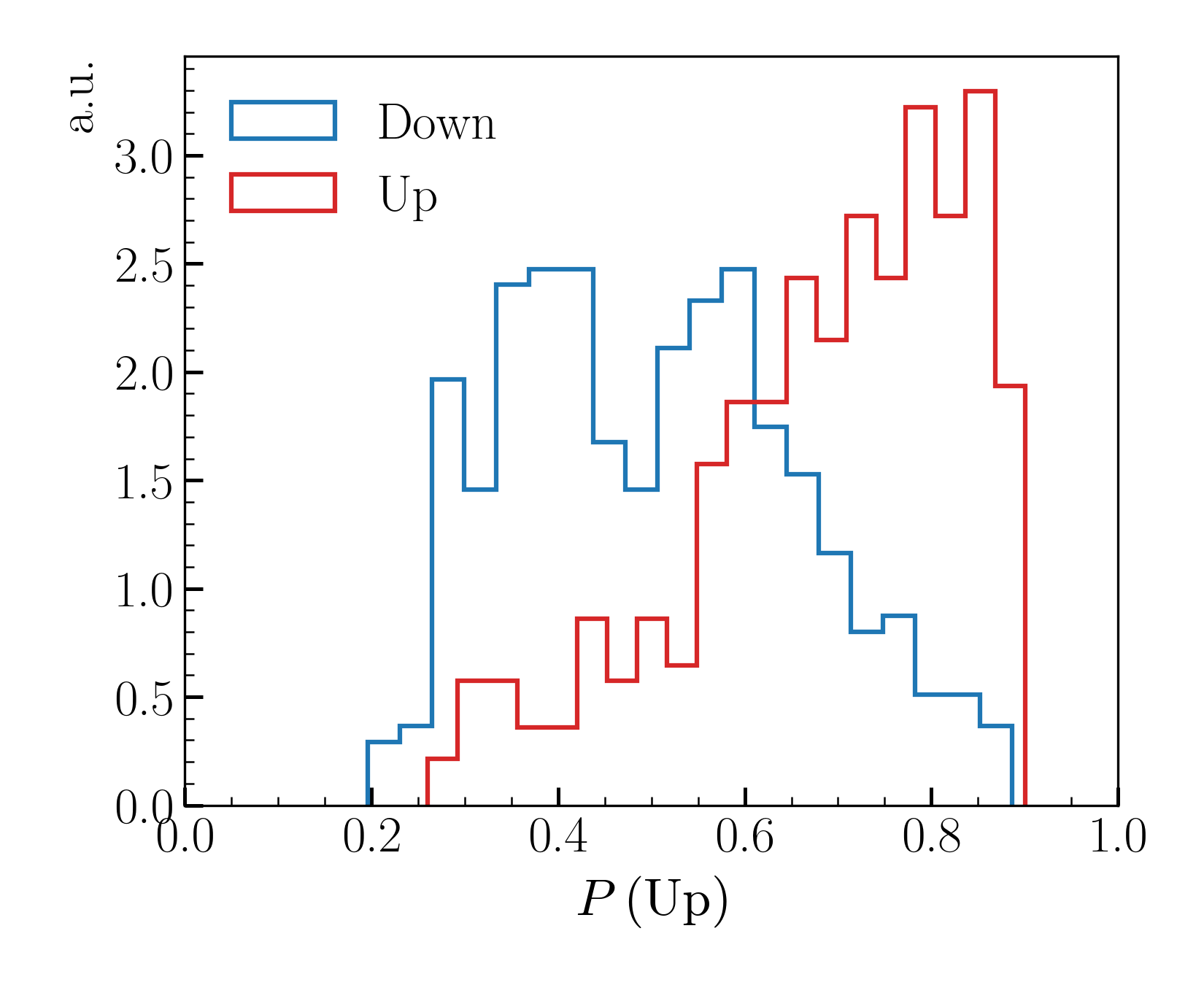}
\end{subfigure}
\hfill
\begin{subfigure}{0.49\linewidth}
    \centering
    \includegraphics[width=\linewidth]{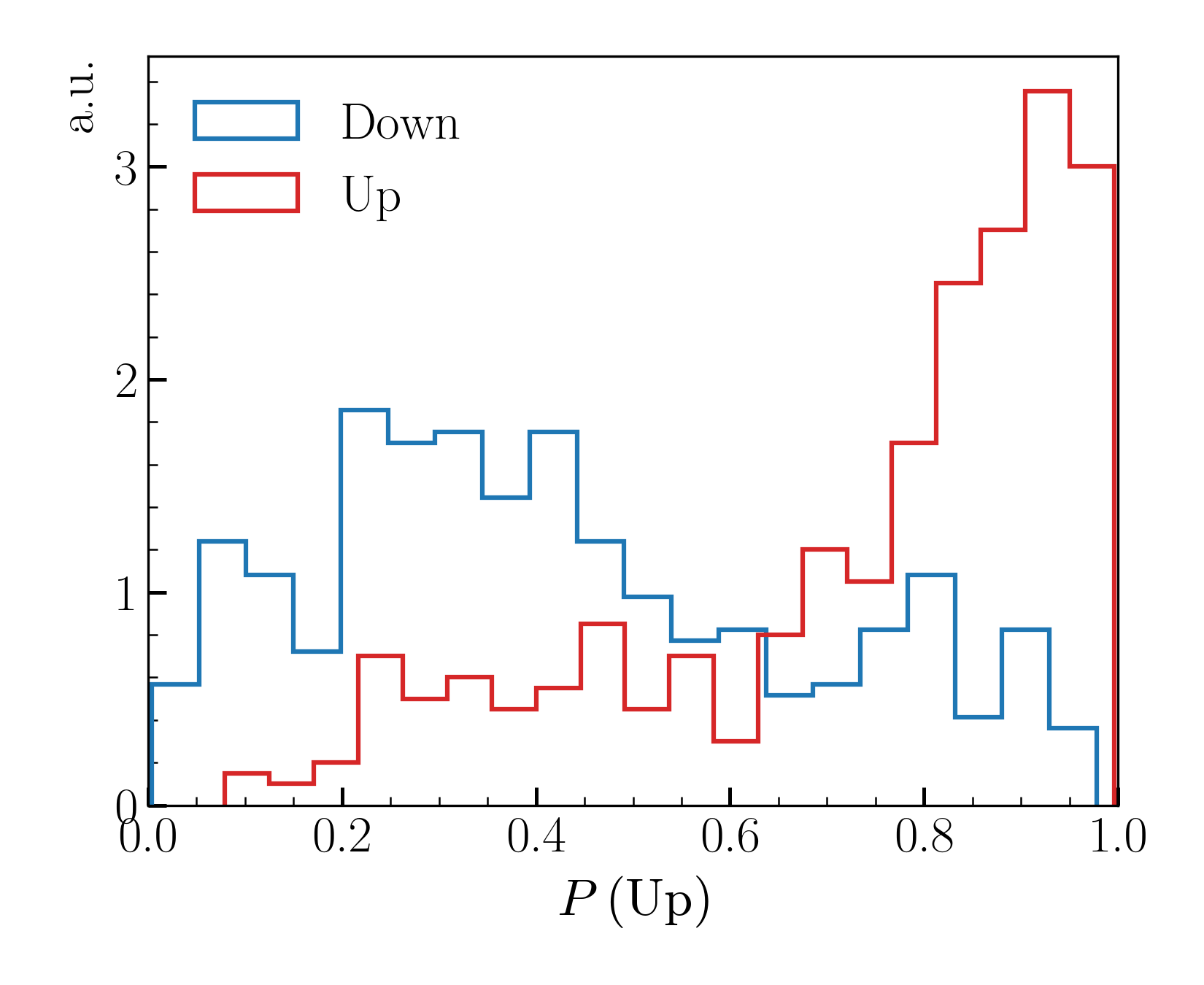}
\end{subfigure}

\caption{QGNN-UD and PFN discrimination histograms on evaluation datasets. (Left) QGNN discriminator output. (Right) PFN discriminator output.}
\label{fig:ud_disc_plots}

\end{figure}

The results of the study are shown in Figure \ref{fig:ud_roc_sic} where the performance of the QGNN-UD, PFN, $Q_{\kappa=0.3}$, and QGNN-Kin are compared. The full set of AUC scores are listed in Table \ref{tab:ud_quc_scores}. We observe that the PFN, jet charge observable, and QGNN-UD model achieve comparable performance within the quoted uncertainties.
Each of the three observables reaches a maximum significance improvement of approximately $1.4$, indicating nontrivial discrimination between the two flavors of jets. To demonstrate the importance of particle ID and charge to the classification, we test the QGNN-Kin model using only particle three-momenta and observe that it does not perform better than random chance. Figure \ref{fig:ud_disc_plots} compares the discriminative power of the QGNN-UD and the PFN models for a given seed, and both models are able to separate both classes.

Purely kinematic observables such as the transverse momentum provide little discrimination power beyond random chance. This is consistent with QCD expectations. Since $u$ and $d$ quarks have the same color representation and negligible masses at the hard-scattering scale, their parton showers and subsequent hadronization produce very similar jet kinematics, making purely kinematic observables only weakly sensitive to the initiating flavor. By contrast, their different electric charges are reflected in the charges and identities of the resulting hadrons. Observables sensitive to particle charge or identity therefore retain residual flavor information that survives hadronization and can distinguish the initiating quark flavor.

\begin{table}[b]
\centering
\caption{Up vs down jet discrimination performance for different models/jet observables over 10 different seeds. The QGNN-UD model is found to perform competitively with each approach. The kinematic QGNN (QGNN-Kin) without particle identification (PID) or charge struggles to separate the two classes of jets.}
\label{tab:ud_quc_scores}
\begin{tabular}{|l|l|c|}
\hline
\textbf{Model} & \textbf{Description} & \textbf{AUC} \\
\hline
\noalign{\vskip 2pt}
\hline
\textbf{PFN} & PFN on all particle features & \textbf{0.801 $\pm$ 0.009} \\
$Q_{\kappa=0.3}$ & Jet charge ($\kappa=0.3$) & 0.800 $\pm$ 0.010 \\
QGNN-UD & QGNN on all particle features & 0.794 $\pm$ 0.009 \\
\hline
$Q_{\kappa=0.5}$ & Jet charge ($\kappa=0.5$) & 0.784 $\pm$ 0.011 \\
$Q_{\kappa=1.0}$ & Jet charge ($\kappa=1.0$) & 0.730 $\pm$ 0.010 \\
QGNN-Kin & QGNN on particle three-momenta only & 0.512 $\pm$ 0.004 \\
\hline
\end{tabular}
\end{table}

\section{Results: Noisy Simulation and Quantum Hardware}\label{sec:hardware}

\begin{table}[b]
\centering
\begin{tabular}{|l|c|c|}
\hline
\textbf{Parameter} & \textbf{Quark-Gluon} & \textbf{Up-Down} \\
\hline
\noalign{\vskip 2pt}
\hline
Loss     & Hinge Loss & Hinge Loss \\
Qubits   & 2 & 4 \\
Layers   & 3 & 2 \\
Jets     & 480 & 500 \\
Features & $z$ & $q,\ |q|\cdot z$ \\
Shots    & 512 & 512 \\
\hline
\end{tabular}
\caption{Hyperparameters for the scaled-down QGNN models executed on the
IBM Heron r2 and IonQ Forte-1 QPUs, where $z = p_{T}/p_{T,\mathrm{Jet}}$ and $q$ is the charge of the particle.}
\label{tab:qpu_hyperparams}
\end{table}

\begin{figure}[t!]
    \centering
    \begin{subfigure}[b]{0.48\textwidth}
        \centering
        \includegraphics[width=\textwidth]{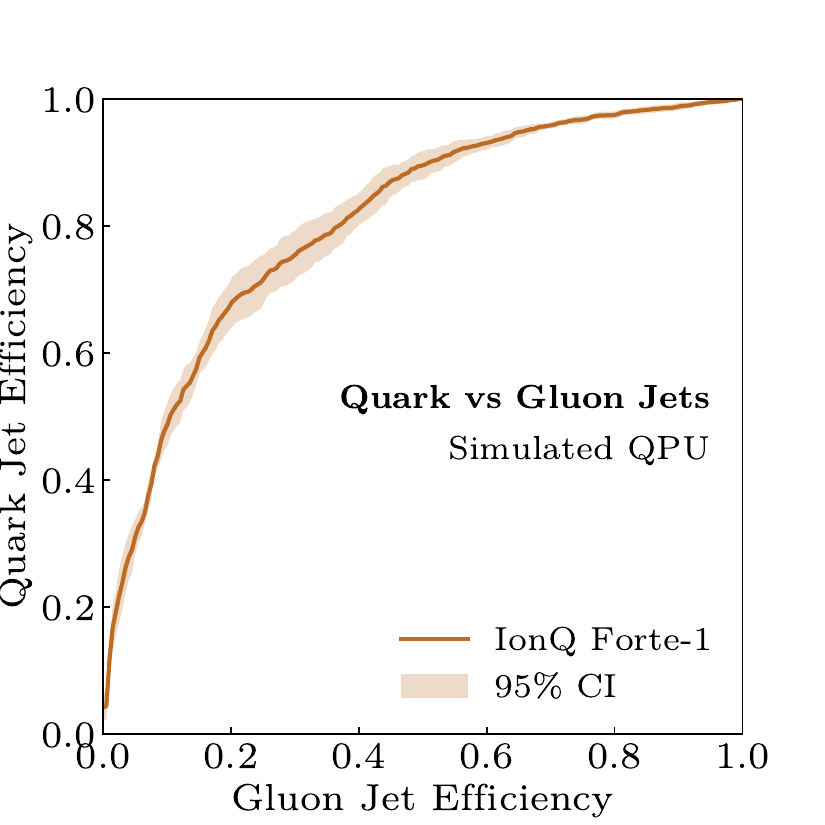}
        \label{fig:sub1_4}
    \end{subfigure}
    \begin{subfigure}[b]{0.48\textwidth}
        \centering
        \includegraphics[width=\textwidth]{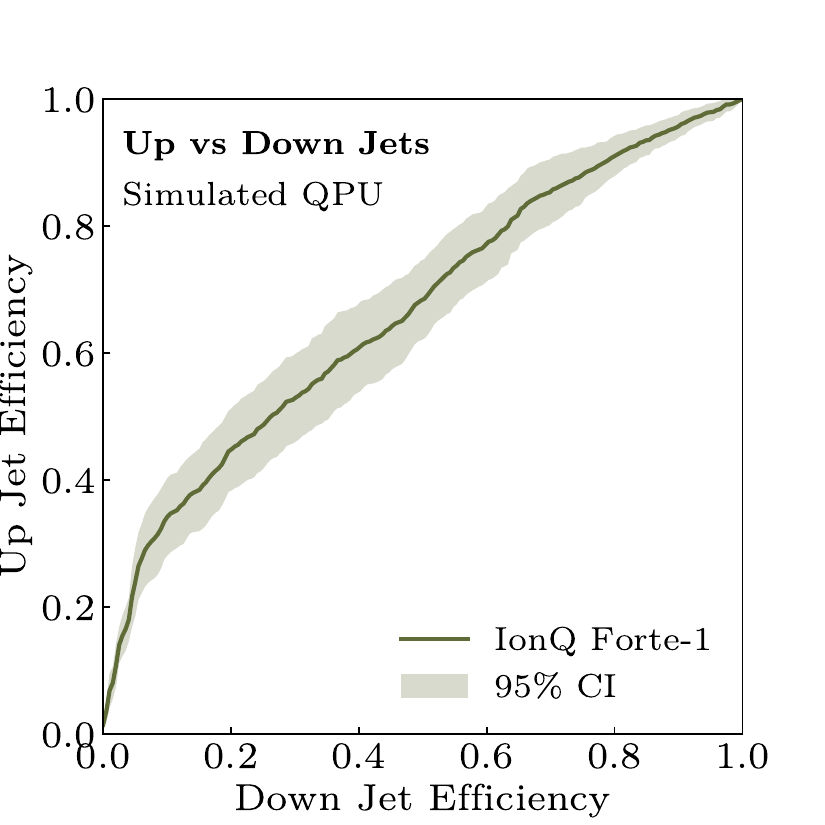}
        \label{fig:sub2_4}
    \end{subfigure}

\caption{Results of simulated noise on the IonQ Forte-1 QPU. We use the same circuit architecture as in the hardware implementation. The quark--gluon and up--down classifications achieve AUCs of $0.807 \pm 0.017$ and $0.665 \pm 0.047$, respectively, averaged over 10 random seeds.}
    \label{fig:ionq_noisy_sim_ci}
\end{figure}

In addition to ideal, noiseless simulation, we demonstrate the QGNN on two architecturally distinct quantum platforms: the superconducting IBM Heron r2 QPU~\cite{Harper:2025bva}, which uses a fixed heavy-hex connectivity lattice, and the trapped-ion IonQ Forte-1 QPU~\cite{Chen:2023erd}, which uses a linear ion chain with all-to-all connectivity. The same ansatz structure is implemented on both platforms, with additional SWAP gates for non-adjacent two-qubit interactions on the IBM device due to its connectivity. The study therefore demonstrates that the QGNN can be trained and evaluated across quantum hardware platforms with distinct physical architectures.

The limited gate fidelity of present-day quantum hardware, together with the substantial sampling cost of on-QPU training, constrains the size and depth of a practically executable circuit, so we deploy scaled-down models with reduced feature inputs relative to the ideal studies of Section~\ref{sec:results_ideal}. The choice of features is guided by domain knowledge behind the construction of hand-crafted QCD observables. For the quark-gluon study, we use only the fractional transverse momentum $z$. For the up-down study, we use $q$ and $|q| \cdot z$, where charge is known to be an essential feature for the discrimination task, and the jet constituents are sorted in descending order of $|q\cdot p_T|$ for truncation. The hyperparameters are summarized in Table~\ref{tab:qpu_hyperparams}. Both models are trained with the same hinge loss used in the ideal studies. Since gradient estimation is notably shot-expensive on hardware, optimization is instead performed with the gradient-free COBYLA optimizer~\cite{Powell:2008udd} limited to a maximum of 30 iterations, with each circuit evaluated using 512 shots.

\begin{figure*}[t]
\centering

\begin{subfigure}{0.4521\textwidth}
    \centering
    \includegraphics[width=\linewidth]{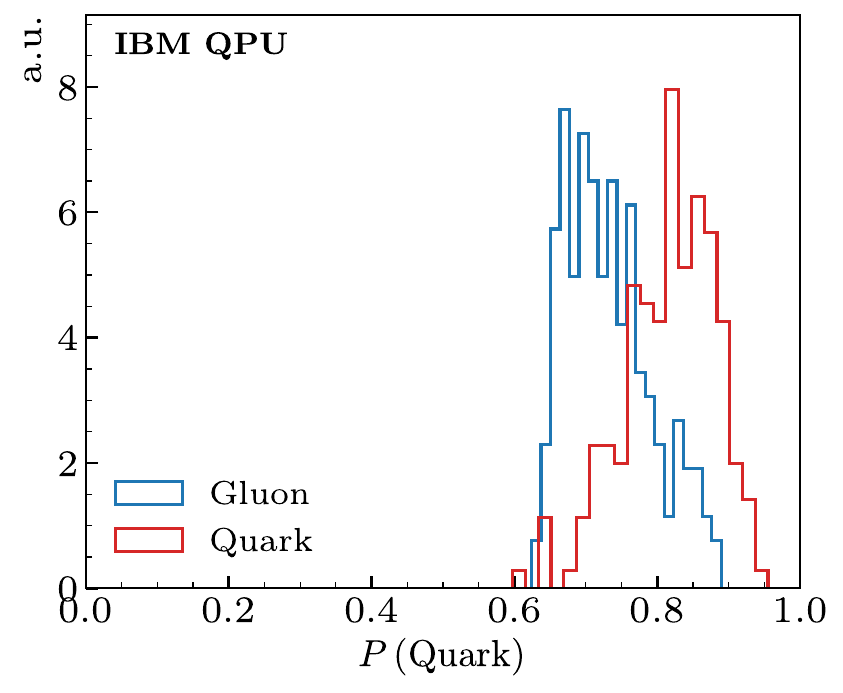}
\end{subfigure}
\hfill
\begin{subfigure}{0.4521\textwidth}
    \centering
    \includegraphics[width=\linewidth]{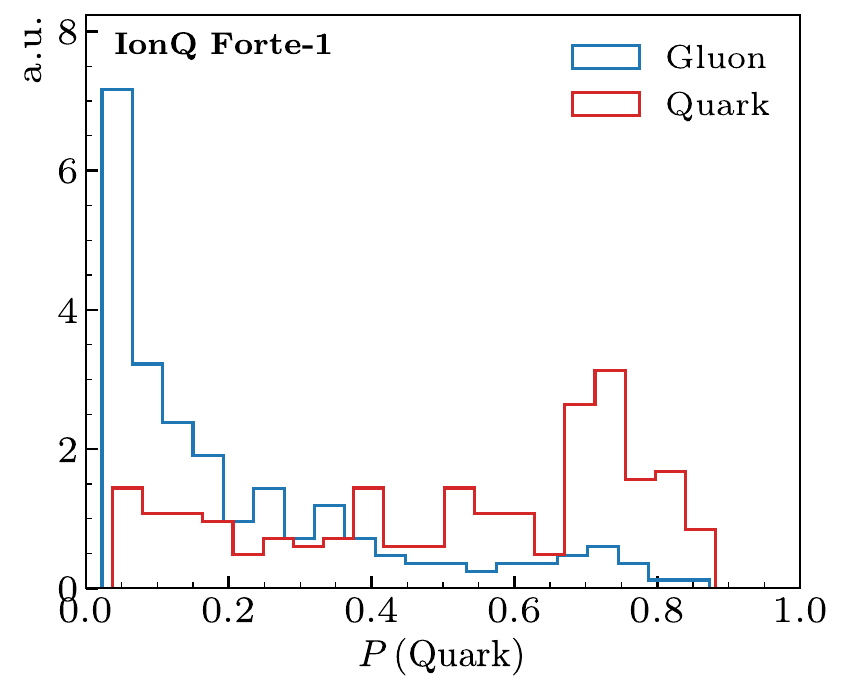}
\end{subfigure}

\vspace{0.5em}

\begin{subfigure}{0.4521\textwidth}
    \centering
    \includegraphics[width=\linewidth]{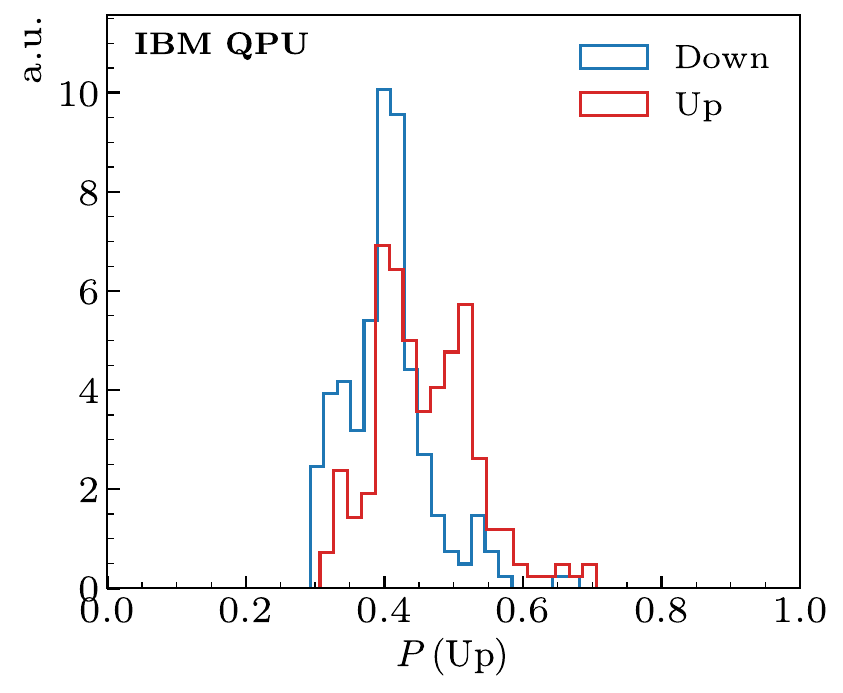}
\end{subfigure}
\hfill
\begin{subfigure}{0.4521\textwidth}
    \centering
    \includegraphics[width=\linewidth]{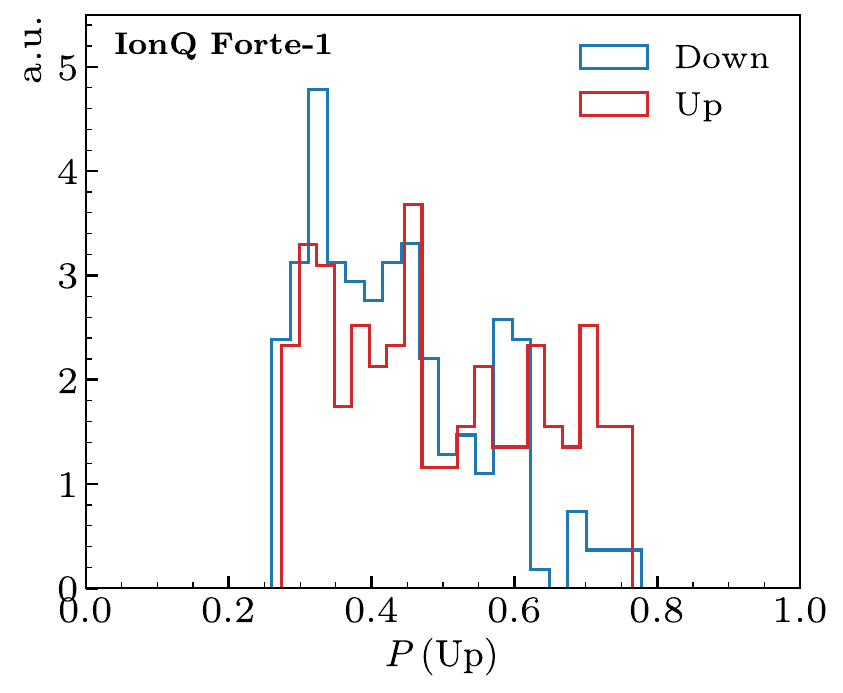}
\end{subfigure}

\caption{Histograms of the classifier outputs for quark-vs-gluon and
$u$- vs.\ $d$-quark discrimination obtained using IBM Heron r2 and IonQ Forte-1
quantum processors. (Top left) Quark-vs-gluon (IBM Heron r2).
(Top right) Quark-vs-gluon (IonQ Forte-1).
(Bottom left) $u$- vs.\ $d$-quark (IBM Heron r2).
(Bottom right) $u$- vs.\ $d$-quark (IonQ Forte-1).}
\label{fig:qpu_histograms}

\end{figure*}


Prior to implementing on quantum hardware, we implement our models with the  simulated noise of a given hardware backend. We use the IonQ Forte-1 noise simulator. We note that the simulator is not an exact representation of hardware noise but allows us to gauge realistic performance. The hyperparameters listed in Table \ref{tab:qpu_hyperparams} are used. The simulated noise model enables multiple independent runs, allowing us to quantify model uncertainty. The resulting mean ROC curves for each task over 10 different seeds, together with their associated uncertainty, are shown in Figure \ref{fig:ionq_noisy_sim_ci}. We observe a wider model variance in the up-down task, indicating that it is more sensitive to stochastic effects during training. It is likely that the reduced dataset, the shallower circuit, and the more challenging nature of the up-down classification task jointly contribute to this observed variance.

\begin{figure*}[t]
    \centering
    \begin{subfigure}{0.32\textwidth}
        \centering
        \includegraphics[width=\linewidth]{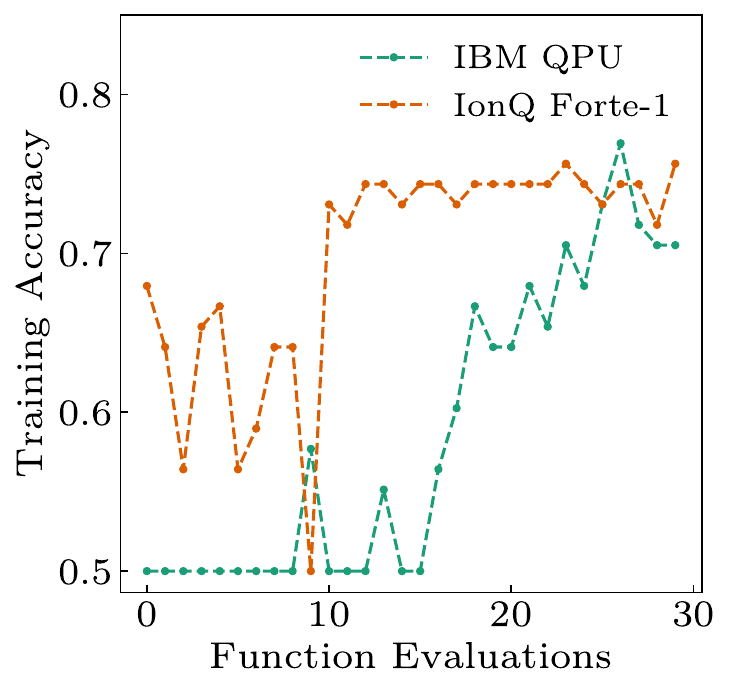}
        \caption{Accuracy}
    \end{subfigure}
    \hfill
    \begin{subfigure}{0.32\textwidth}
        \centering
        \includegraphics[width=\linewidth]{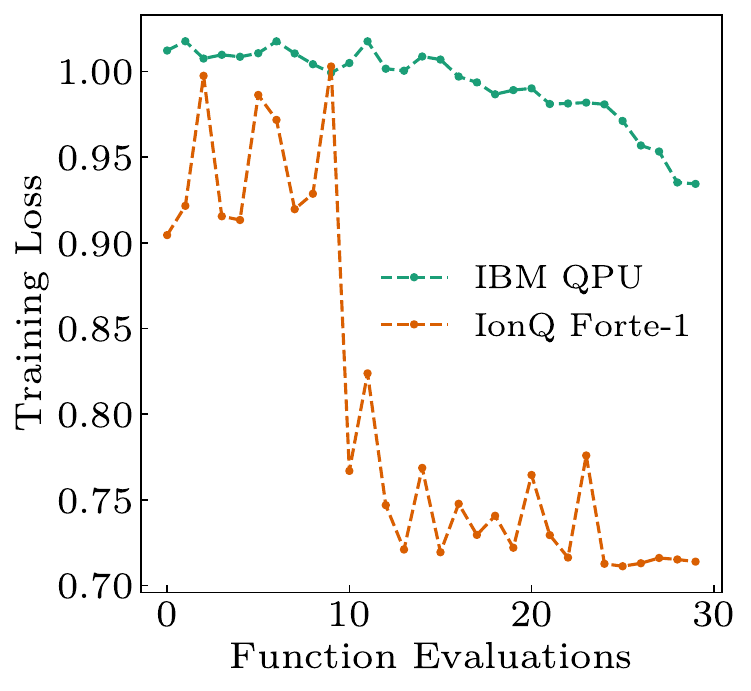}
        \caption{Loss}
    \end{subfigure}
    \hfill
    \begin{subfigure}{0.32\textwidth}
        \centering
        \includegraphics[width=\linewidth]{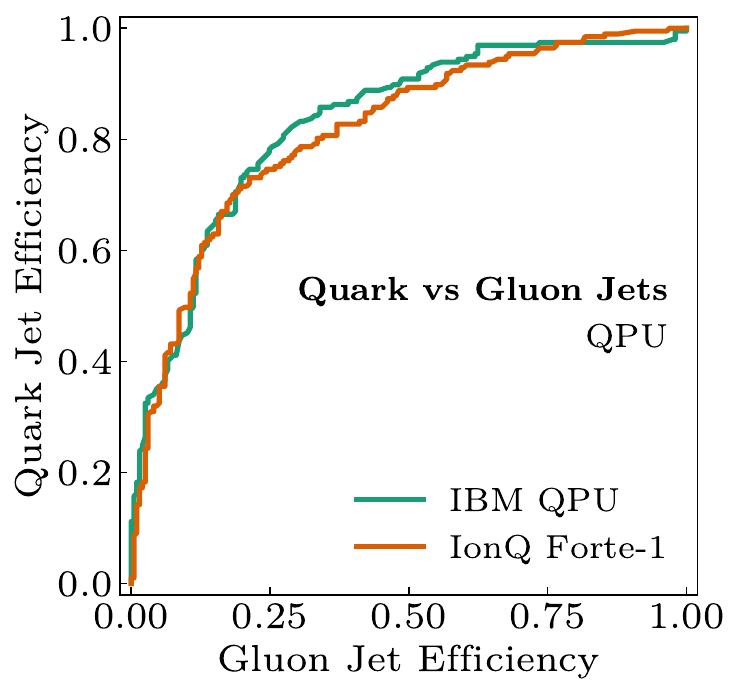}
        \caption{ROC}
    \end{subfigure}

    \caption{Comparison of quark-vs-gluon jet discrimination performance on
    IBM Heron r2 and IonQ Forte-1 QPUs. The panels show the classification accuracy,
    training loss, and ROC curves, respectively.}
    \label{fig:qg_stacked}
\end{figure*}

\begin{figure*}[t]
    \centering
    \begin{subfigure}{0.32\textwidth}
        \centering
        \includegraphics[width=\linewidth]{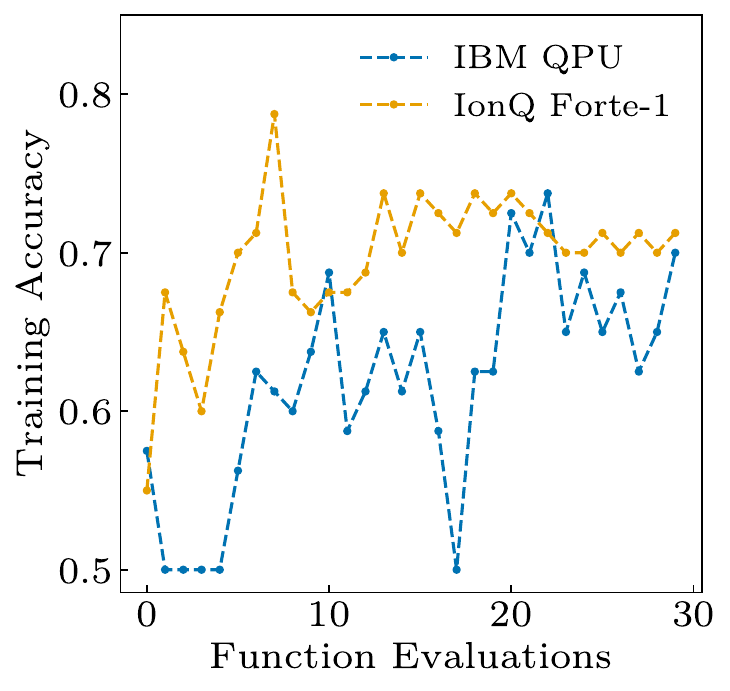}
        \caption{Accuracy}
    \end{subfigure}
    \hfill
    \begin{subfigure}{0.32\textwidth}
        \centering
        \includegraphics[width=\linewidth]{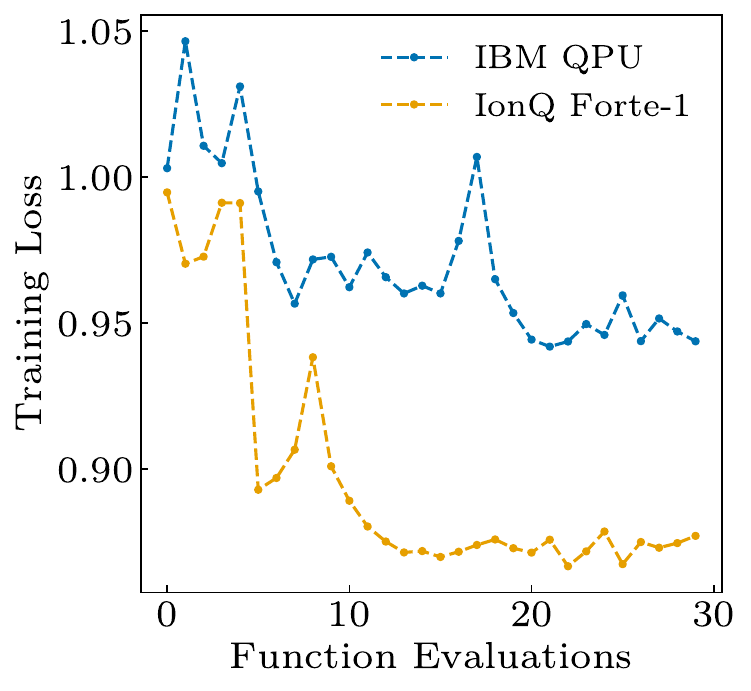}
        \caption{Loss}
    \end{subfigure}
    \hfill
    \begin{subfigure}{0.32\textwidth}
        \centering
        \includegraphics[width=\linewidth]{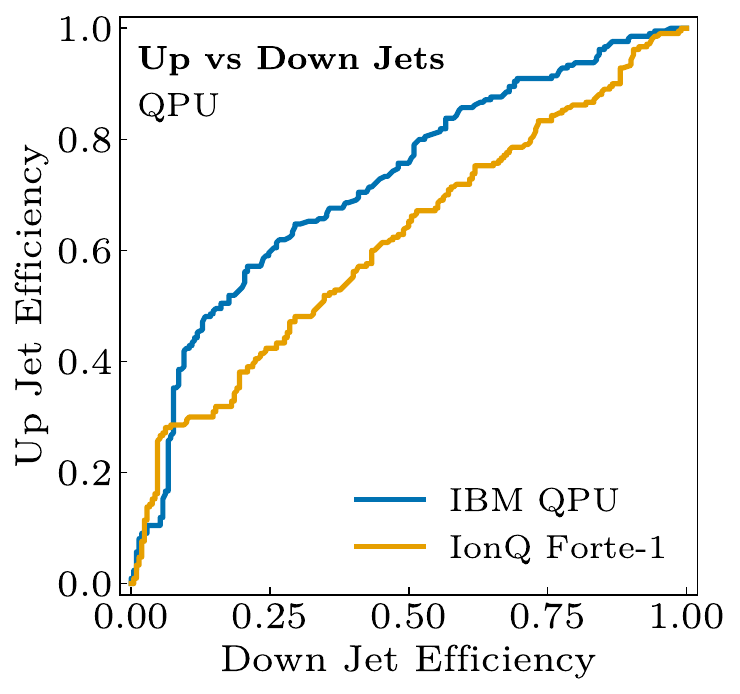}
        \caption{ROC}
    \end{subfigure}

    \caption{Comparison of $u$- vs.\ $d$-quark discrimination performance on
    IBM Heron r2 and IonQ Forte-1 QPUs. The panels show the classification accuracy,
    training loss, and ROC curves, respectively.}
    \label{fig:ud_stacked}
\end{figure*}

With the noisy simulation study providing a baseline for expected hardware performance, we next train and evaluate the models directly on the IBM Heron r2 and IonQ Forte-1 QPUs. Figure~\ref{fig:qpu_histograms} shows the discrimination histograms for both tasks on both devices, demonstrating a strong discrimination power between quarks and gluons but relatively weaker discrimination power between up and down jets. Figures~\ref{fig:qg_stacked} and
\ref{fig:ud_stacked} present the training accuracy, training loss, and ROC curves for the quark-gluon and flavor studies respectively. For the quark-gluon task, the hardware models achieve AUCs of $0.830$ (IBM Heron r2) and $0.815$ (IonQ Forte-1), closely tracking the ideal-simulation result of Section~\ref{sec:results_ideal} despite the drastic reduction in qubit count and circuit depth. This indicates that the compact two-qubit model retains the kinematic discrimination power identified in the interpretability analysis and that it executes robustly on current hardware.

For the flavor tagging task, the hardware models achieve AUCs of $0.721$ (IBM Heron r2) and $0.619$ (IonQ Forte-1). The up-versus-down task relies more strongly on constituent-level charge information, so aggressive truncation can remove useful discriminating information. In our particle-to-qubit encoding, retaining more constituents requires more qubits, while current hardware constraints necessitate a smaller and shallower circuit. Consequently, some degradation in discrimination performance relative to the ideal model is observed. The disparity between the IBM and IonQ results may be attributed to random fluctuations during training over the relatively small hardware dataset. We expect that a larger model, closer to the ideal configuration of Section~\ref{sec:results_ideal}, would yield stronger flavor discrimination, at correspondingly greater computational cost on hardware.

\section{Model Interpretability}\label{sec:int}

Interpretable machine learning methods have recently gathered interest within the particle physics community~\cite{Wetzel:2025uhj, Ngairangbam:2023cps, taskinKnowledgeIntegrationPhysicsinformed2026, Bogatskiy:2023nnw}. The key motivation is to shift from a treatment of neural networks and associated architectures as black-boxes producing opaque predictions to explainable tools. It seeks to address questions of whether a given machine learning model is learning meaningful physical correlations and which features or observables drive its predictions. A deep understanding of why any classifier works is, as a result, useful for developing more robust methodologies and searching for new physics. 

In this section, we investigate the discriminating power of the QGNN to demonstrate its ability to learn meaningful differences between each class of jets considered. We motivate our analysis as follows:
\begin{itemize}
    \item \textit{A priori}, we expect certain particle-level features to drive the prediction whereas other such features are not as impactful. We determine whether the model is able to identify these features. 
    \item If the model exploits information captured by established jet observables, we expect its output to exhibit statistical dependence on these observables.
\end{itemize}

We address the first point of the analysis by comparing the weight vectors $\vec{w}$ across all layers. Since the per-particle features can have different characteristic scales, the magnitudes of their raw encoding weights are not directly comparable with each other. To permit comparison among the weight vectors, we scale each weight by the standard deviation of the corresponding feature ($\sigma_f$). Therefore, for an important feature $f$ to the model's predictions, we expect $|w_f\sigma_f|$ to be relatively large.
Conversely, for a relatively unimportant feature, we expect that $|w_f\sigma_f|$ to be close to zero. To characterize cross-layer importance of a particular feature $f$, we compute the associated $L_1, L_2$ norms as follows:

\[
I^{(L_1)}_f
=
\sum_{l=1}^{L}\left|w_f^l \sigma_f\right|,
\qquad
I^{(L_2)}_f
=
\sqrt{\sum_{l=1}^{L}\left(w_f^l\sigma_f\right)^2}
\]
where $L$ is the total number of ansatz layers in the associated QGNN model.

\begin{figure}[t]
    \centering
    \includegraphics[width=\linewidth]{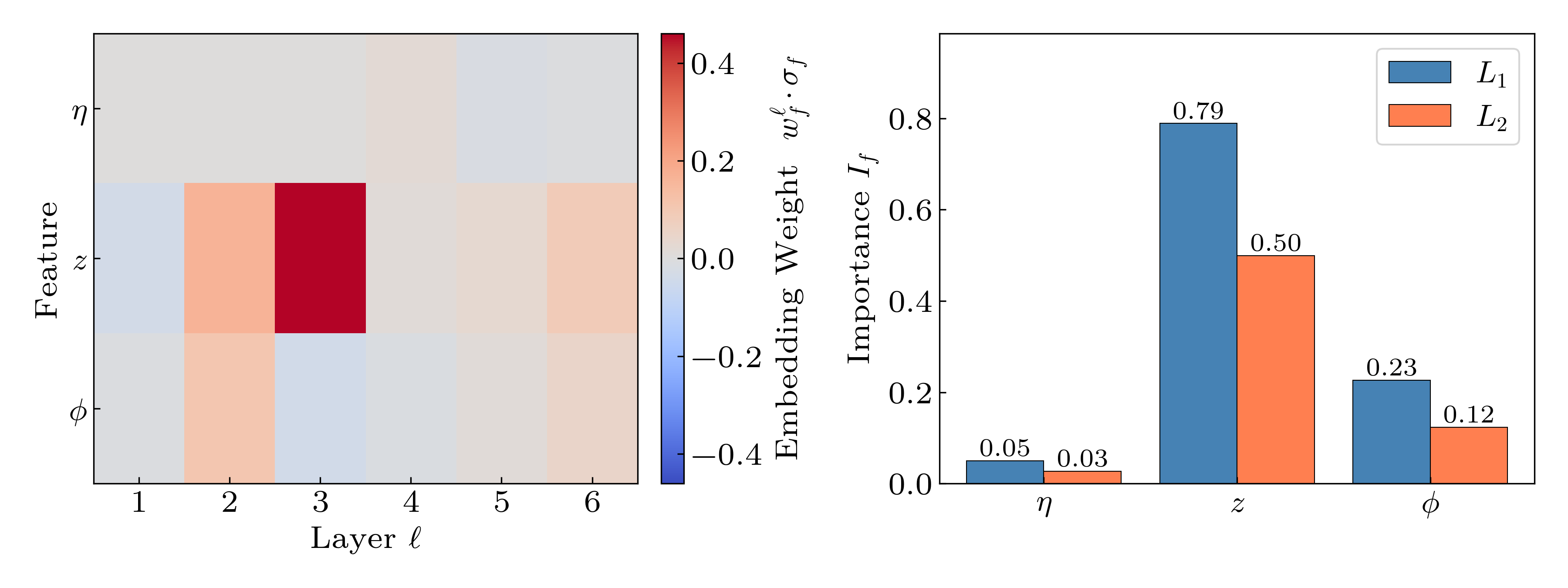}
    \caption{$L_1$, $L_2$ norms computed on each feature's cross-layer weights in the QGNN-QG model. The average over 10 unique seeds is taken.}
    \label{fig:qg_norms}
\end{figure}

For the QGNN-QG model, Figure \ref{fig:qg_norms} shows the computed $L_1, L_2$ norms across the PQC. We observe that the fractional transverse momentum $z$ has the highest aggregated encoding weight compared to $\eta, \phi$. For the angular features, $|w_{\eta}\sigma_\eta| \approx 0$ across all layers and ${I}_\phi^{(L_1)}, {I}_\phi^{(L_2)}$ are relatively small, suggesting that the model most strongly relies on $z$ for its predictions.    

\begin{figure}[t]
    \centering
    \includegraphics[width=\textwidth]{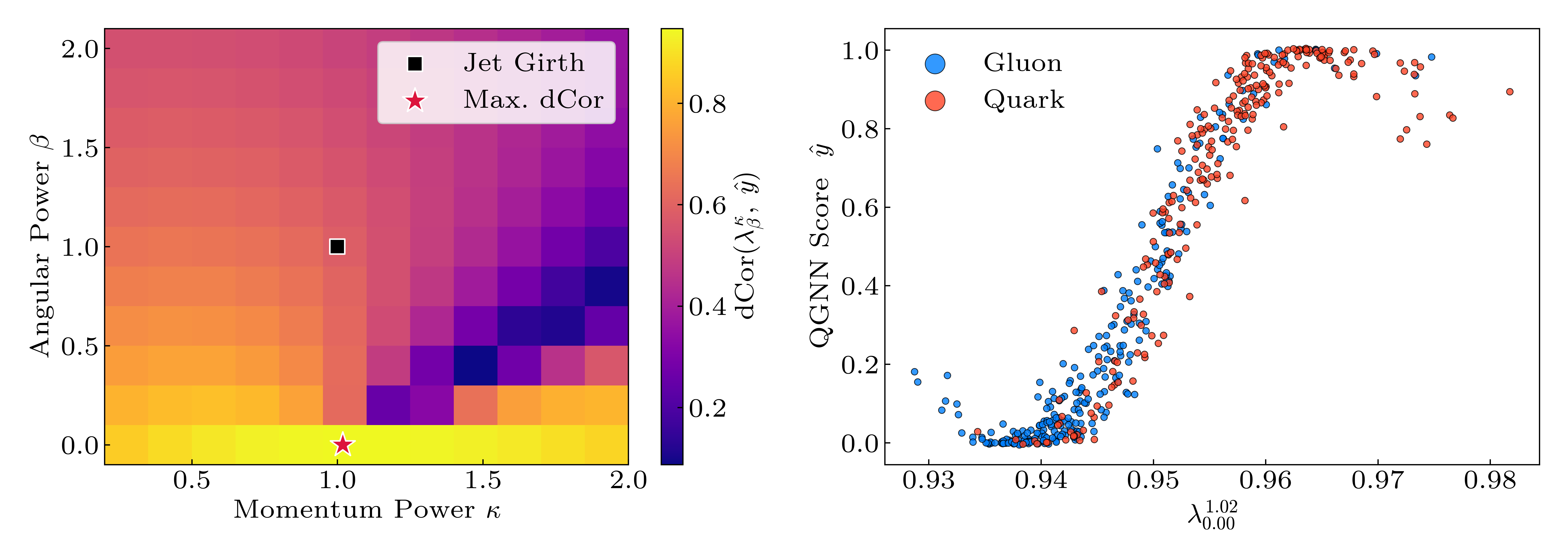}
    \caption{(Left) dCor($\lambda_{\beta}^{\kappa}, \hat{y})$ for angularities scanned across various values of $\beta, \kappa$ with respect to the model output $\hat{y}$. We find the strongest correlation for $\beta \approx 0.00$ and $\kappa \approx 1.02$ with dCor = $0.958 \pm 0.007$. (Right) Correlation of $\hat{y}$ with $\lambda_{0.00}^{1.02}$.}

    \label{fig:qg_dcor_angs}
\end{figure}

To address the second point of our interpretability analysis, we parametrize a set of observables relevant to each classification task and determine whether the model outputs correlate with them. In particular, we consider the family of jet angularity observables~\mbox{\cite{Berger:2003iw,Larkoski:2014pca,Kang:2018qra}}.

\[
\mathrm{\mathbf{Generalized \ Angularities:}} \quad \lambda_\beta ^\kappa =  \sum_{i\in J} z_i ^\kappa \left( \frac{R_i}{R} \right)^\beta,
\]
where $R$ is the jet radius and $R_i$ is the angular fraction with respect to the jet axis. They are quark-gluon jet discriminators because they probe the angular and energetic structure of jets. To determine the angularity most closely learned by the model, we compute the distance correlation (dCor)~\cite{Szekely2007} between the QGNN-QG score $(\hat{y})$ and the angularity ($\lambda_\beta^\kappa$) for different $\kappa, \beta$. Here, we consider the output scores of a single seed. Figure \ref{fig:qg_dcor_angs} shows the results of this test. We find that 
\[
    \lambda_{0.00}^{1.02} = \sum_{i\in J} z_i ^{1.02}
\]
has the highest dCor with respect to $\hat{y}$ and therefore correlates most strongly with the model outputs. We remark that the dCor-maximizing exponent is close to $1.00$ with $\beta \approx 0$. Since $\lambda_0^1$ corresponds to the fraction of jet transverse momentum retained after truncation, this indicates that the QGNN score is strongly sensitive to the retained momentum fraction. This is consistent with Figure~\ref{fig:removed_pt}, where the fixed-$N$ truncation leads to different momentum losses for quark and gluon jets due to their different constituent multiplicities. In particular, gluon jets typically have higher constituent multiplicity and softer fragmentation, leading to a larger momentum loss under the truncation. The retained momentum fraction can therefore serve as a proxy for these underlying quark-gluon differences in the truncated jet representation.

\begin{figure}[t]
    \centering
    \includegraphics[width=\linewidth]{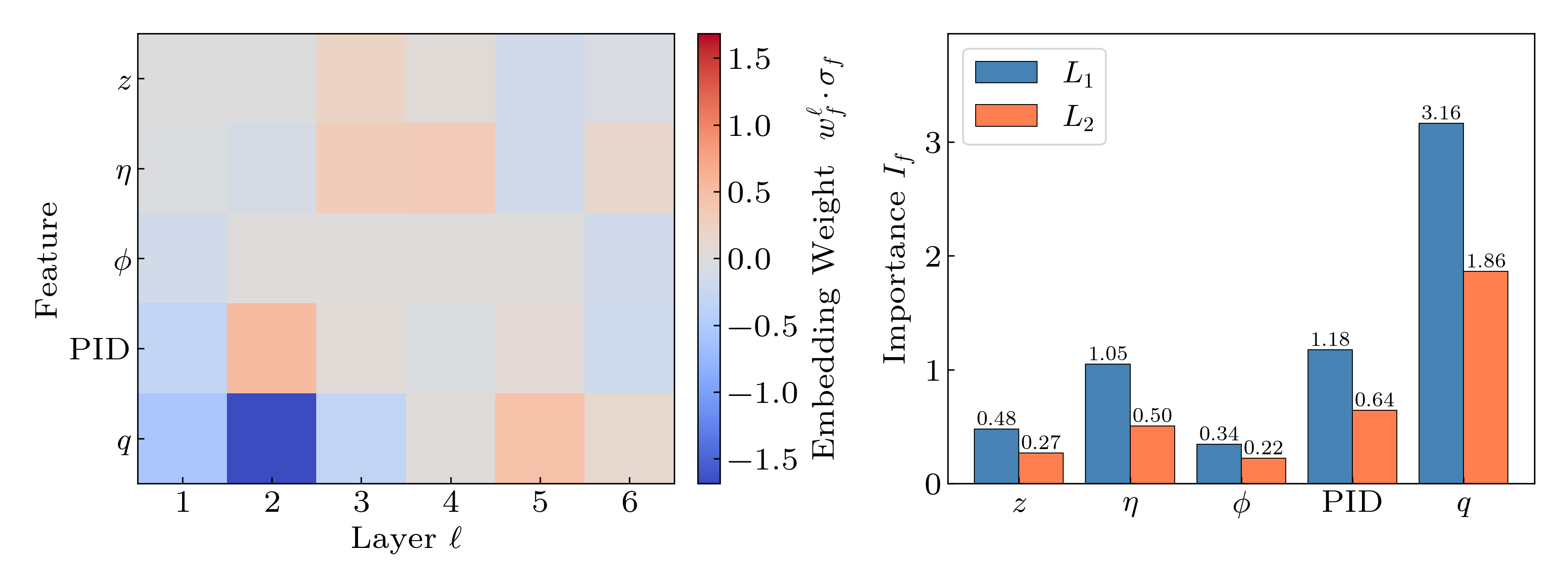}
    \caption{$L_1$, $L_2$ norms computed on each feature's cross-layer weights in the QGNN-UD model. The average over 10 unique seeds is taken.}
    \label{fig:ud_norms}
\end{figure}

\begin{figure}[b]
    \centering
    \includegraphics[width=\textwidth]{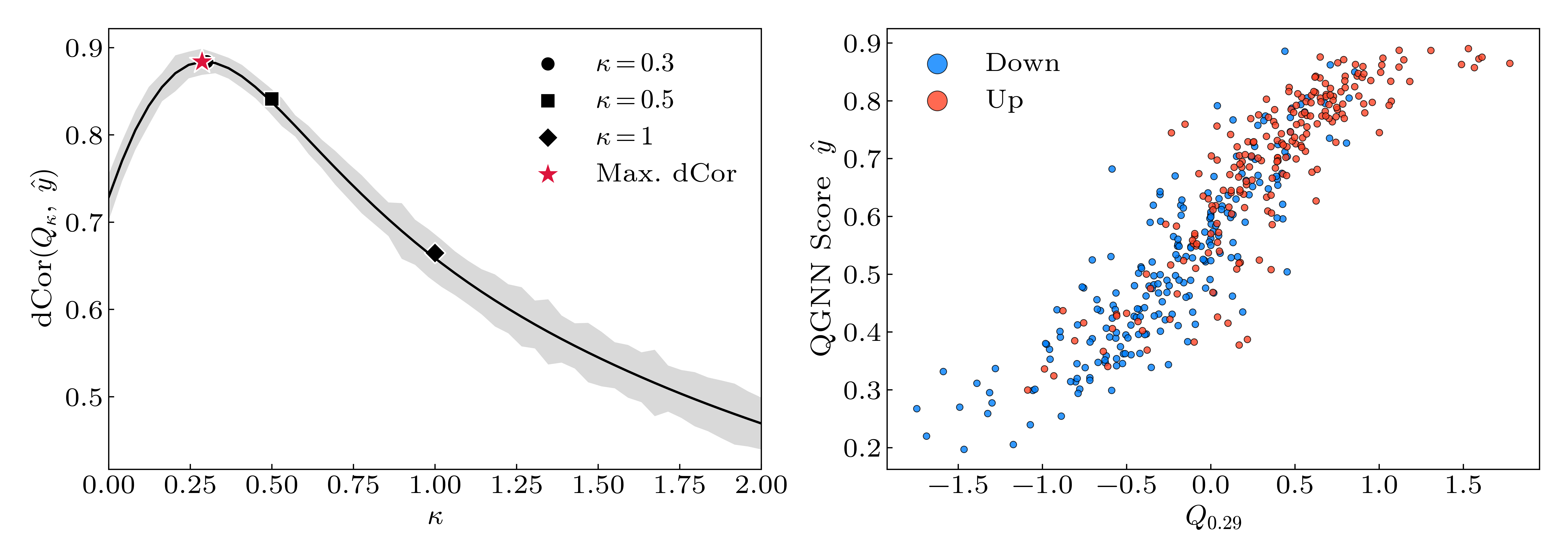}
    \caption{(Left) dCor$(Q_{\kappa}, \hat{y})$ for jet charge scanned across various values of $\kappa$ with respect to the model output $\hat{y}$. We find the strongest correlation for $\kappa \approx 0.29$ with dCor $ = 0.884 \pm 0.029$. (Right) Correlation of $\hat{y}$ with $Q_{0.29}$.}
    \label{fig:ud_dcor_jcs}
\end{figure}

Figure \ref{fig:ud_norms} shows the $L_1, L_2$ norms across the PQC for the QGNN-UD model. As expected, the charge $q$ carries the most importance, followed by PID and $\eta$. Relatively less importance is attributed to $z$ and $\phi$, unlike the QGNN-QG model. Additionally, this explains why the QGNN-Kin model severely underperforms without this full set of particle-level features, especially charge, as detailed in Figure \ref{fig:ud_roc_sic} and Table \ref{tab:ud_quc_scores}. 

To characterize the QGNN-UD learning with respect to hand-crafted QCD observables, we correlate the output of a single trained model with the jet charge observable, for which we scan over $\kappa$. Figure~\ref{fig:ud_dcor_jcs} shows the dCor between the jet charge and QGNN-UD scores for various values of $\kappa$, as well as the correlation at the dCor-maximizing value of $\kappa$. We find a strong correlation between the two observables for $\kappa \approx 0.29$. The strong statistical dependence on jet charge suggests that the QGNN exploits charge-weighted momentum information similar to that encoded by jet charge. However, this dependence alone does not imply that jet charge fully accounts for the model predictions; instead, it suggests that the model output reflects, at least in part, information captured by jet charge.

\section{Conclusions and Outlook}\label{sec:conclusion}

We have presented a permutation-invariant Quantum Graph Neural Network for jet classification, respecting the permutation symmetry of jet particle cloud representations. The model is then applied to essential jet tagging tasks for the LHC and future EIC: quark-gluon and up-down jet discrimination, where the latter task is expected to be more challenging.

In comparison to classical machine learning algorithms such as PFN and hand-crafted observables such as jet charge and jet girth, we observe that the idealized, permutation-symmetry-preserving quantum model achieves comparable performance on both tasks. Additionally, the implementation of reduced models on IBM and IonQ quantum hardware yields promising results despite hardware noise. The interpretability study shows that the quantum model leverages specific particle-level features for its predictions. In the quark-gluon study, we find that the QGNN-QG relies strongly on the per-particle fractional momentum $z$ and its output is strongly correlated with the $\lambda^{1.02}_{0.00}$ angularity. On the other hand, in the flavor study, the QGNN-UD leverages charge $q$, particle identification (PID), and $\eta$ per-particle features, and the QGNN-UD output is strongly correlated with the jet charge $Q_{\kappa=0.29}$ observable. These results demonstrate that the quantum model is capturing physically relevant information from the particle cloud.

The insight we highlight is that shallow symmetry-invariant quantum models can offer strong predictions when implemented on hardware backends using small datasets. To the best of our knowledge, this is the first implementation of direct training of a quantum model on QPUs for jet tagging. As a demonstration of the capabilities of Noisy Intermediate-Scale Quantum (NISQ) hardware, it is promising that one can achieve solid performance and retain meaningful discrimination with shallow quantum models under noisy environments. Further, this highlights the ability of current QPUs to complete non-trivial learning tasks, specifically in jet tagging. Beyond the performance of the models, we have demonstrated that symmetry-preserving models are able to leverage particle-level features for distinct classification tasks. Additionally, correlations between the model outputs and established QCD observables provide insight into the physical information used by the model.

As quantum hardware continues to mature through increased qubit counts, improved gate fidelities, more effective error correction, and hardware-efficient quantum algorithms, increasingly expressive and deeper symmetry-preserving QML models will become more accessible. While the practical deployment of QML for large-scale HEP analyses remains a longer-term objective, the results presented here establish that meaningful jet classification can already be performed on current quantum processors. This provides a concrete benchmark for future hardware and algorithmic developments, and jet tagging is a realistic testbed for evaluating the future improvements of quantum machine learning in high-energy physics.

\section*{Data and Code Availability}

The datasets used in this study are publicly available from the sources cited in the text. The code used to train and evaluate the quantum and classical models, as well as to produce the results presented in this work, will be made available upon publication.

\acknowledgments

We would like to thank Luke Sellers and Diego Padilla for helpful discussions. BJ was supported by the Physics and Astronomy REU program at UCLA. BJ and ZK are supported by the National Science Foundation under Grant No.~PHY-2515057. JY is supported by the National Quantum Laboratory (QLab) at the University of Maryland. We also thank QLab and IBM Quantum for providing access to computing resources.

\FloatBarrier

\FloatBarrier

\bibliographystyle{JHEP}
\bibliography{references.bib}

@article{Ikeda:2025bjb,
    author = "Ikeda, Kazuki and Kang, Zhong-Bo and Kharzeev, Dmitri E. and Qian, Wenyang",
    title = "{Quantum simulation of real-time current correlators and DIS-inspired observables in the Schwinger model}",
    eprint = "2512.18062",
    archivePrefix = "arXiv",
    primaryClass = "hep-ph",
    doi = "10.1007/JHEP07(2026)242",
    journal = "JHEP",
    volume = "07",
    pages = "242",
    year = "2026"
}

@article{Kang:2020fka,
    author = "Kang, Zhong-Bo and Liu, Xiaohui and Mantry, Sonny and Shao, Ding Yu",
    title = "{Jet Charge: A Flavor Prism for Spin Asymmetries at the EIC}",
    eprint = "2008.00655",
    archivePrefix = "arXiv",
    primaryClass = "hep-ph",
    doi = "10.1103/PhysRevLett.125.242003",
    journal = "Phys. Rev. Lett.",
    volume = "125",
    pages = "242003",
    year = "2020"
}

@article{Kang:2021ryr,
    author = "Kang, Zhong-Bo and Liu, Xiaohui and Mantry, Sonny and Spraker, M. C. and Wilson, Tyler",
    title = "{Dynamic Jet Charge}",
    eprint = "2101.04304",
    archivePrefix = "arXiv",
    primaryClass = "hep-ph",
    doi = "10.1103/PhysRevD.103.074028",
    journal = "Phys. Rev. D",
    volume = "103",
    number = "7",
    pages = "074028",
    year = "2021"
}

@article{Fraser:2018ieu,
    author = "Fraser, Katherine and Schwartz, Matthew D.",
    title = "{Jet Charge and Machine Learning}",
    eprint = "1803.08066",
    archivePrefix = "arXiv",
    primaryClass = "hep-ph",
    doi = "10.1007/JHEP10(2018)093",
    journal = "JHEP",
    volume = "10",
    pages = "093",
    year = "2018"
}

@article{Larkoski:2017jix,
    author = "Larkoski, Andrew J. and Moult, Ian and Nachman, Benjamin",
    title = "{Jet Substructure at the Large Hadron Collider: A Review of Recent Advances in Theory and Machine Learning}",
    eprint = "1709.04464",
    archivePrefix = "arXiv",
    primaryClass = "hep-ph",
    doi = "10.1016/j.physrep.2019.11.001",
    journal = "Phys. Rept.",
    volume = "841",
    pages = "1--63",
    year = "2020"
}

@article{AbdulKhalek:2021gbh,
    author = "Abdul Khalek, R. and others",
    title = "{Science Requirements and Detector Concepts for the Electron-Ion Collider}: {EIC Yellow Report}",
    eprint = "2103.05419",
    archivePrefix = "arXiv",
    primaryClass = "physics.ins-det",
    reportNumber = "BNL-220990-2021-FORE, JLAB-PHY-21-3198, LA-UR-21-20953",
    doi = "10.1016/j.nuclphysa.2022.122447",
    journal = "Nucl. Phys. A",
    volume = "1026",
    pages = "122447",
    year = "2022"
}

@article{Kang:2018qra,
    author = "Kang, Zhong-Bo and Lee, Kyle and Ringer, Felix",
    title = "{Jet angularity measurements for single inclusive jet production}",
    eprint = "1801.00790",
    archivePrefix = "arXiv",
    primaryClass = "hep-ph",
    doi = "10.1007/JHEP04(2018)110",
    journal = "JHEP",
    volume = "04",
    pages = "110",
    year = "2018"
}

@article{Napolitano:2026gge,
    author = "Napolitano, Fabrizio and Della Penna, Luca and Tedeschi, Tommaso and Fan{\`o}, Livio",
    title = "{Lund Plane to Bloch (LP$^{2}$B) encoding for object and polarization tagging with quantum jet substructure}",
    eprint = "2604.18613",
    archivePrefix = "arXiv",
    primaryClass = "quant-ph",
    doi = "10.1140/epjc/s10052-026-16142-9",
    journal = "Eur. Phys. J. C",
    volume = "86",
    number = "7",
    pages = "909",
    year = "2026"
}

@article{Terashi:2020wfi,
    author = "Terashi, Koji and Kaneda, Michiru and Kishimoto, Tomoe and Saito, Masahiko and Sawada, Ryu and Tanaka, Junichi",
    title = "{Event Classification with Quantum Machine Learning in High-Energy Physics}",
    eprint = "2002.09935",
    archivePrefix = "arXiv",
    primaryClass = "physics.comp-ph",
    doi = "10.1007/s41781-020-00047-7",
    journal = "Comput. Softw. Big Sci.",
    volume = "5",
    number = "1",
    pages = "2",
    year = "2021"
}

@article{Harper:2025bva,
    author = "Harper, Robin and Lain{\'e}, Constance and Hockings, Evan T. and McLauchlan, Campbell and Nixon, Georgia M. and Brown, Benjamin J. and Bartlett, Stephen D.",
    title = "{Characterising the failure mechanisms of error-corrected quantum logic gates}",
    eprint = "2504.07258",
    archivePrefix = "arXiv",
    primaryClass = "quant-ph",
    doi = "10.1038/s41467-026-71773-6",
    journal = "Nature Commun.",
    volume = "17",
    number = "1",
    pages = "5039",
    year = "2026"
}

@article{Chen:2023erd,
    author = "Chen, Jwo-Sy and others",
    title = "{Benchmarking a trapped-ion quantum computer with 30 qubits}",
    eprint = "2308.05071",
    archivePrefix = "arXiv",
    primaryClass = "quant-ph",
    doi = "10.22331/q-2024-11-07-1516",
    journal = "Quantum",
    volume = "8",
    pages = "1516",
    year = "2024"
}

@article{Andrews:2021ejw,
    author = "Andrews, Michael and others",
    title = "{End-to-end jet classification of boosted top quarks with the CMS open data}",
    eprint = "2104.14659",
    archivePrefix = "arXiv",
    primaryClass = "physics.data-an",
    doi = "10.1051/epjconf/202125104030",
    journal = "EPJ Web Conf.",
    volume = "251",
    pages = "04030",
    year = "2021"
}

@article{Accardi:2012qut,
    author = "Accardi, A. and others",
    editor = "Deshpande, A. and Meziani, Z. E. and Qiu, J. W.",
    title = "{Electron Ion Collider: The Next QCD Frontier}: {Understanding the glue that binds us all}",
    eprint = "1212.1701",
    archivePrefix = "arXiv",
    primaryClass = "nucl-ex",
    reportNumber = "BNL-98815-2012-JA, JLAB-PHY-12-1652",
    doi = "10.1140/epja/i2016-16268-9",
    journal = "Eur. Phys. J. A",
    volume = "52",
    number = "9",
    pages = "268",
    year = "2016"
}

@article{Arratia:2020azl,
    author = "Arratia, Miguel and Furletova, Yulia and Hobbs, T. J. and Olness, Fredrick and Sekula, Stephen J.",
    title = "{Charm jets as a probe for strangeness at the future Electron-Ion Collider}",
    eprint = "2006.12520",
    archivePrefix = "arXiv",
    primaryClass = "hep-ph",
    reportNumber = "JLAB-PHY-20-3205, SMU-HEP-20-05",
    doi = "10.1103/PhysRevD.103.074023",
    journal = "Phys. Rev. D",
    volume = "103",
    number = "7",
    pages = "074023",
    year = "2021"
}

@article{Wu:2022tnc,
    author = "Wu, Sau Lan and others",
    title = "{Application of Quantum Machine Learning to High Energy Physics Analysis at LHC Using Quantum Computer Simulators and Quantum Computer Hardware}",
    reportNumber = "FERMILAB-CONF-22-331-DI-QIS",
    doi = "10.22323/1.398.0842",
    journal = "PoS",
    volume = "EPS-HEP2021",
    pages = "842",
    year = "2022"
}

@article{Singh:2024dyh,
    author = "Singh, Utkarsh and Goldberg, Aaron Z. and Heshami, Khabat",
    title = "{Coherent feed-forward quantum neural network}",
    eprint = "2402.00653",
    archivePrefix = "arXiv",
    primaryClass = "quant-ph",
    doi = "10.1007/s42484-024-00222-8",
    journal = "Quantum Machine Intelligence",
    volume = "6",
    number = "2",
    pages = "89",
    year = "2024"
}

@article{Wach:2023ufx, 
    author = "Wach, Noah L. and Rudolph, Manuel S. and Jendrzejewski, Fred and Schmitt, Sebastian",
    title = "{Data re-uploading with a single qudit}",
    eprint = "2302.13932",
    archivePrefix = "arXiv",
    primaryClass = "quant-ph",
    doi = "10.1007/s42484-023-00125-0",
    journal = "Quantum Machine Intelligence",
    volume = "5",
    number = "2",
    pages = "36",
    year = "2023"
}

@article{Thaler:2010tr, 
    author = "Thaler, Jesse and Van Tilburg, Ken",
    title = "{Identifying Boosted Objects with N-subjettiness}",
    eprint = "1011.2268",
    archivePrefix = "arXiv",
    primaryClass = "hep-ph",
    reportNumber = "MIT-CTP-4191",
    doi = "10.1007/JHEP03(2011)015",
    journal = "JHEP",
    volume = "03",
    pages = "015",
    year = "2011"
}

@article{Powell:2008udd, 
    author = "Powell, M. J. D.",
    title = "{Direct search algorithms for optimization calculations}",
    doi = "10.1017/S0962492900002841",
    journal = "Acta Numer.",
    volume = "7",
    pages = "287--336",
    year = "2008"
}

@inproceedings{Kingma:2014vow, 
    author = "Kingma, Diederik P. and Ba, Jimmy",
    title = "{Adam: A Method for Stochastic Optimization}",
    booktitle = "{International Conference on Learning Representations}",
    eprint = "1412.6980",
    archivePrefix = "arXiv",
    primaryClass = "cs.LG",
    month = "12",
    year = "2014"
}

@article{Larkoski:2019nwj,
    author = "Larkoski, Andrew J. and Metodiev, Eric M.",
    title = "{A Theory of Quark vs. Gluon Discrimination}",
    eprint = "1906.01639",
    archivePrefix = "arXiv",
    primaryClass = "hep-ph",
    reportNumber = "MIT-CTP 5049",
    doi = "10.1007/JHEP10(2019)014",
    journal = "JHEP",
    volume = "10",
    pages = "014",
    year = "2019"
}

@article{Subba:2022czw,
    author = "Subba, Amir and Singh, Ritesh K.",
    title = "{Role of polarizations and spin-spin correlations of W's in e-e+{\textrightarrow}W-W+ at s=250{\,}{\,}GeV to probe anomalous W-W+Z/{\ensuremath{\gamma}} couplings}",
    eprint = "2212.12973",
    archivePrefix = "arXiv",
    primaryClass = "hep-ph",
    doi = "10.1103/PhysRevD.107.073004",
    journal = "Phys. Rev. D",
    volume = "107",
    number = "7",
    pages = "073004",
    year = "2023"
}

@article{FerreiradeLima:2016gcz,
    author = "Ferreira de Lima, Danilo and Petrov, Petar and Soper, Davison and Spannowsky, Michael",
    title = "{Quark-Gluon tagging with Shower Deconstruction: Unearthing dark matter and Higgs couplings}",
    eprint = "1607.06031",
    archivePrefix = "arXiv",
    primaryClass = "hep-ph",
    reportNumber = "DCPT-16-140, IPPP-16-70",
    doi = "10.1103/PhysRevD.95.034001",
    journal = "Phys. Rev. D",
    volume = "95",
    number = "3",
    pages = "034001",
    year = "2017"
}

@article{Szekely2007,
  author    = {Sz{\'e}kely, G{\'a}bor J. and Rizzo, Maria L. and Bakirov, Nail K.},
  title     = {Measuring and Testing Dependence by Correlation of Distances},
  journal   = {The Annals of Statistics},
  year      = {2007},
  volume    = {35},
  number    = {6},
  pages     = {2769--2794},
  doi       = {10.1214/009053607000000505},
  publisher = {Institute of Mathematical Statistics}
}

@software{deepmind2020jax,
  author = {James Bradbury and Roy Frostig and Peter Hawkins and Matthew James Johnson and Yash Katariya and Chris Leary and Dougal Maclaurin and George Necula and Adam Paszke and Jake Vander{P}las and Skye Wanderman-{M}ilne and Qiao Zhang},
  title = {{JAX}: composable transformations of {P}ython+{N}um{P}y programs},
  url = {http://github.com/jax-ml/jax},
  version = {0.3.13},
  year = {2018},
}

@article{Perez-Salinas:2019pjx, 
    author = "P{\'e}rez-Salinas, Adri{\'a}n and Cervera-Lierta, Alba and Gil-Fuster, Elies and Latorre, Jos{\'e} I.",
    title = "{Data re-uploading for a universal quantum classifier}",
    eprint = "1907.02085",
    archivePrefix = "arXiv",
    primaryClass = "quant-ph",
    doi = "10.22331/q-2020-02-06-226",
    journal = "Quantum",
    volume = "4",
    pages = "226",
    year = "2020"
}

@article{Cacciari:2011ma, 
    author = "Cacciari, Matteo and Salam, Gavin P. and Soyez, Gregory",
    title = "{FastJet User Manual}",
    eprint = "1111.6097",
    archivePrefix = "arXiv",
    primaryClass = "hep-ph",
    reportNumber = "CERN-PH-TH-2011-297",
    doi = "10.1140/epjc/s10052-012-1896-2",
    journal = "Eur. Phys. J. C",
    volume = "72",
    pages = "1896",
    year = "2012"
}

@article{Qu:2022mxj,
    author = "Qu, Huilin and Li, Congqiao and Qian, Sitian",
    title = "{Particle Transformer for Jet Tagging}",
    eprint = "2202.03772",
    archivePrefix = "arXiv",
    primaryClass = "hep-ph",
    month = "2",
    year = "2022"
}

@article{Field:1977fa,
    author = "Field, R. D. and Feynman, R. P.",
    editor = "Brown, L. M.",
    title = "{A Parametrization of the Properties of Quark Jets}",
    reportNumber = "CALT-68-618",
    doi = "10.1016/0550-3213(78)90015-9",
    journal = "Nucl. Phys. B",
    volume = "136",
    pages = "1",
    year = "1978"
}

@article{Lee:2022kdn,
    author = "Lee, Kyle and Mulligan, James and P{\l}osko{\'n}, Mateusz and Ringer, Felix and Yuan, Feng",
    title = "{Machine learning-based jet and event classification at the Electron-Ion Collider with applications to hadron structure and spin physics}",
    eprint = "2210.06450",
    archivePrefix = "arXiv",
    primaryClass = "hep-ph",
    reportNumber = "JLAB-THY-22-3739, MIT-CTP 5473, YITP-SB-2022-34",
    doi = "10.1007/JHEP03(2023)085",
    journal = "JHEP",
    volume = "03",
    pages = "085",
    year = "2023"
}

@article{Mernyei:2021krm,
    author = "Mernyei, P{\'e}ter and Meichanetzidis, Konstantinos and Ceylan, {\.I}smail {\.I}lkan",
    title = "{Equivariant Quantum Graph Circuits}",
    eprint = "2112.05261",
    archivePrefix = "arXiv",
    primaryClass = "cs.LG",
    month = "12",
    year = "2021"
}

@article{Schatzki:2022tfq,
    author = "Schatzki, Louis and Larocca, Martin and Nguyen, Quynh T. and Sauvage, Frederic and Cerezo, M.",
    title = "{Theoretical guarantees for permutation-equivariant quantum neural networks}",
    eprint = "2210.09974",
    archivePrefix = "arXiv",
    primaryClass = "quant-ph",
    reportNumber = "LA-UR-22-29899",
    doi = "10.1038/s41534-024-00804-1",
    journal = "npj Quantum Inf.",
    volume = "10",
    number = "1",
    pages = "12",
    year = "2024"
}

@article{Chen:2024rna,
    author = "Chen, Yi-An and Chen, Kai-Feng",
    title = "{Jet discrimination with a quantum complete graph neural network}",
    eprint = "2403.04990",
    archivePrefix = "arXiv",
    primaryClass = "hep-ph",
    doi = "10.1103/PhysRevD.111.016020",
    journal = "Phys. Rev. D",
    volume = "111",
    number = "1",
    pages = "016020",
    year = "2025"
}

@article{Elhag:2024xfw,
    author = "Elhag, Hala and Hartung, Tobias and Jansen, Karl and Nagano, Lento and Pirina, Giorgio Menicagli and Di Tucci, Alice",
    title = "{Quantum convolutional neural networks for jet images classification}",
    eprint = "2408.08701",
    archivePrefix = "arXiv",
    primaryClass = "quant-ph",
    month = "8",
    year = "2024"
}

@misc{komiskePythia8QuarkGluon2019,
	title = {Pythia8 {Quark} and {Gluon} {Jets} for {Energy} {Flow}},
	copyright = {Creative Commons Attribution 4.0 International, Open Access},
	url = {https://zenodo.org/record/3164691},
	doi = {10.5281/ZENODO.3164691},
	urldate = {2026-02-03},
	publisher = {Zenodo},
	author = {Komiske, Patrick and Metodiev, Eric and Thaler, Jesse},
	month = may,
	year = {2019},
}

@misc{leePYTHIA6DatasetMachine2023,
	title = {{PYTHIA6} {Dataset}: {Machine} learning-based jet and event classification at the {Electron}-{Ion} {Collider} with applications to hadron structure and spin physics},
	shorttitle = {{PYTHIA6} {Dataset}},
	url = {https://zenodo.org/records/7538810},
	doi = {10.5281/zenodo.7538810},
	urldate = {2026-02-03},
	publisher = {Zenodo},
	author = {Lee, Kyle and Mulligan, James and Ploskon, Mateusz and Ringer, Felix and Yuan, Feng},
	month = jan,
	year = {2023},
}

@article{Komiske:2018cqr,
    author = "Komiske, Patrick T. and Metodiev, Eric M. and Thaler, Jesse",
    title = "{Energy Flow Networks: Deep Sets for Particle Jets}",
    eprint = "1810.05165",
    archivePrefix = "arXiv",
    primaryClass = "hep-ph",
    reportNumber = "MIT-CTP 5064",
    doi = "10.1007/JHEP01(2019)121",
    journal = "JHEP",
    volume = "01",
    pages = "121",
    year = "2019"
}

@article{Qu:2019gqs,
    author = "Qu, Huilin and Gouskos, Loukas",
    title = "{ParticleNet: Jet Tagging via Particle Clouds}",
    eprint = "1902.08570",
    archivePrefix = "arXiv",
    primaryClass = "hep-ph",
    doi = "10.1103/PhysRevD.101.056019",
    journal = "Phys. Rev. D",
    volume = "101",
    number = "5",
    pages = "056019",
    year = "2020"
}

@article{Li:2026ydk,
    author = "Li, Ting and Liu, Shanglong and Xu, GuangZhi and Xie, Peizhong",
    title = "{Quantum complete graph self-attention network for particle flow classification}",
    doi = "10.1088/2632-2153/ae32dd",
    journal = "Mach. Learn. Sci. Tech.",
    volume = "7",
    number = "1",
    pages = "015007",
    year = "2026"
}

@article{Jahin:2024zss,
    author = "Jahin, Md Abrar and Masud, Md. Akmol and Suva, Md Wahiduzzaman and Mridha, M. F. and Dey, Nilanjan",
    title = "{Lorentz-Equivariant Quantum Graph Neural Network for High-Energy Physics}",
    eprint = "2411.01641",
    archivePrefix = "arXiv",
    primaryClass = "cs.LG",
    doi = "10.1109/TAI.2025.3554461",
    month = "11",
    year = "2024"
}

@article{Bal:2025ydm,
    author = "Bal, Aritra and Klute, Markus and Maier, Benedikt and Oughton, Melik and Pezone, Eric and Spannowsky, Michael",
    title = "{One particle - one qubit: Particle physics data encoding for quantum machine learning}",
    eprint = "2502.17301",
    archivePrefix = "arXiv",
    primaryClass = "hep-ph",
    reportNumber = "IPPP/25/11",
    doi = "10.1103/l8y2-87vq",
    journal = "Phys. Rev. D",
    volume = "112",
    number = "7",
    pages = "076004",
    year = "2025"
}

@article{Jahin:2024wjw,
    author = "Jahin, Md Abrar and Masud, Md. Akmol and Mridha, M. F. and Dey, Nilanjan and Aung, Zeyar",
    title = "{Quantum Rationale-Aware Graph Contrastive Learning for Jet Discrimination}",
    eprint = "2411.01642",
    archivePrefix = "arXiv",
    primaryClass = "cs.LG",
    month = "11",
    year = "2024"
}

@article{Cerezo:2020jpv,
    author = "Cerezo, M. and others",
    title = "{Variational quantum algorithms}",
    eprint = "2012.09265",
    archivePrefix = "arXiv",
    primaryClass = "quant-ph",
    reportNumber = "LA-UR-20-30142",
    doi = "10.1038/s42254-021-00348-9",
    journal = "Nature Rev. Phys.",
    volume = "3",
    number = "9",
    pages = "625--644",
    year = "2021"
}

@article{Biamonte:2016ugo,
    author = "Biamonte, Jacob and Wittek, Peter and Pancotti, Nicola and Rebentrost, Patrick and Wiebe, Nathan and Lloyd, Seth",
    title = "{Quantum machine learning}",
    eprint = "1611.09347",
    archivePrefix = "arXiv",
    primaryClass = "quant-ph",
    doi = "10.1038/nature23474",
    journal = "Nature",
    volume = "549",
    number = "7671",
    pages = "195--202",
    year = "2017"
}

@article{Zhang:2025raf,
    author = "Zhang, Shan-Liang and Wang, Enke and Wang, Xin-Nian and Xing, Hongxi",
    title = "{Unraveling the neutron skin thickness through jet charge in deep inelastic scattering}",
    eprint = "2506.10694",
    archivePrefix = "arXiv",
    primaryClass = "hep-ph",
    month = "6",
    year = "2025"
}

@article{Li:2019dre,
    author = "Li, Hai Tao and Vitev, Ivan",
    title = "{Jet charge modification in dense QCD matter}",
    eprint = "1908.06979",
    archivePrefix = "arXiv",
    primaryClass = "hep-ph",
    reportNumber = "LA-UR-19-30442",
    doi = "10.1103/PhysRevD.101.076020",
    journal = "Phys. Rev. D",
    volume = "101",
    pages = "076020",
    year = "2020"
}

@article{ATLAS:2015rlw,
    author = "Aad, Georges and others",
    collaboration = "ATLAS",
    title = "{Measurement of jet charge in dijet events from $\sqrt{s}$=8  TeV pp collisions with the ATLAS detector}",
    eprint = "1509.05190",
    archivePrefix = "arXiv",
    primaryClass = "hep-ex",
    reportNumber = "CERN-PH-EP-2015-207",
    doi = "10.1103/PhysRevD.93.052003",
    journal = "Phys. Rev. D",
    volume = "93",
    number = "5",
    pages = "052003",
    year = "2016"
}

@article{Gallicchio:2011xq,
    author = "Gallicchio, Jason and Schwartz, Matthew D.",
    title = "{Quark and Gluon Tagging at the LHC}",
    eprint = "1106.3076",
    archivePrefix = "arXiv",
    primaryClass = "hep-ph",
    doi = "10.1103/PhysRevLett.107.172001",
    journal = "Phys. Rev. Lett.",
    volume = "107",
    pages = "172001",
    year = "2011"
}

@article{Farrell:2024fit,
    author = "Farrell, Roland C. and Illa, Marc and Ciavarella, Anthony N. and Savage, Martin J.",
    title = "{Quantum simulations of hadron dynamics in the Schwinger model using 112 qubits}",
    eprint = "2401.08044",
    archivePrefix = "arXiv",
    primaryClass = "quant-ph",
    reportNumber = "IQuS@UW-21-069, NT@UW-24-1",
    doi = "10.1103/PhysRevD.109.114510",
    journal = "Phys. Rev. D",
    volume = "109",
    number = "11",
    pages = "114510",
    year = "2024"
}

@article{Farrell:2023fgd,
    author = "Farrell, Roland C. and Illa, Marc and Ciavarella, Anthony N. and Savage, Martin J.",
    title = "{Scalable Circuits for Preparing Ground States on Digital Quantum Computers: The Schwinger Model Vacuum on 100 Qubits}",
    eprint = "2308.04481",
    archivePrefix = "arXiv",
    primaryClass = "quant-ph",
    reportNumber = "IQuS@UW-21-060, NT@UW-23-13",
    doi = "10.1103/PRXQuantum.5.020315",
    journal = "PRX Quantum",
    volume = "5",
    number = "2",
    pages = "020315",
    year = "2024"
}

@article{Klco:2018kyo,
    author = "Klco, N. and Dumitrescu, E. F. and McCaskey, A. J. and Morris, T. D. and Pooser, R. C. and Sanz, M. and Solano, E. and Lougovski, P. and Savage, M. J.",
    title = "{Quantum-classical computation of Schwinger model dynamics using quantum computers}",
    eprint = "1803.03326",
    archivePrefix = "arXiv",
    primaryClass = "quant-ph",
    reportNumber = "INT-PUB-18-013",
    doi = "10.1103/PhysRevA.98.032331",
    journal = "Phys. Rev. A",
    volume = "98",
    number = "3",
    pages = "032331",
    year = "2018"
}

@article{Kokail:2018eiw,
    author = "Kokail, Christian and others",
    title = "{Self-verifying variational quantum simulation of lattice models}",
    eprint = "1810.03421",
    archivePrefix = "arXiv",
    primaryClass = "quant-ph",
    doi = "10.1038/s41586-019-1177-4",
    journal = "Nature",
    volume = "569",
    number = "7756",
    pages = "355--360",
    year = "2019"
}

@article{Wetzel:2025uhj,
    author = "Wetzel, Sebastian Johann and Ha, Seungwoong and Iten, Raban and Klopotek, Miriam and Liu, Ziming",
    title = "{Interpretable Machine Learning in Physics: A Review}",
    eprint = "2503.23616",
    archivePrefix = "arXiv",
    primaryClass = "physics.comp-ph",
    month = "3",
    year = "2025"
}

@article{Zohar:2013zla,
    author = "Zohar, Erez and Cirac, J. Ignacio and Reznik, Benni",
    title = "{Quantum simulations of gauge theories with ultracold atoms: local gauge invariance from angular momentum conservation}",
    eprint = "1303.5040",
    archivePrefix = "arXiv",
    primaryClass = "quant-ph",
    doi = "10.1103/PhysRevA.88.023617",
    journal = "Phys. Rev. A",
    volume = "88",
    pages = "023617",
    year = "2013"
}

@article{Lamm:2019bik,
    author = "Lamm, Henry and Lawrence, Scott and Yamauchi, Yukari",
    title = "{General Methods for Digital Quantum Simulation of Gauge Theories}",
    eprint = "1903.08807",
    archivePrefix = "arXiv",
    primaryClass = "hep-lat",
    doi = "10.1103/PhysRevD.100.034518",
    journal = "Phys. Rev. D",
    volume = "100",
    number = "3",
    pages = "034518",
    year = "2019"
}

@article{Vigl:2026ppx,
    author = "Vigl, Matthias and Hartman, Nicole and Kagan, Michael and Heinrich, Lukas",
    title = "{Neural Scaling Laws for Boosted Jet Tagging}",
    eprint = "2602.15781",
    archivePrefix = "arXiv",
    primaryClass = "hep-ex",
    month = "2",
    year = "2026"
}

@article{Dasgupta:2013ihk,
    author = "Dasgupta, Mrinal and Fregoso, Alessandro and Marzani, Simone and Salam, Gavin P.",
    title = "{Towards an understanding of jet substructure}",
    eprint = "1307.0007",
    archivePrefix = "arXiv",
    primaryClass = "hep-ph",
    reportNumber = "CERN-PH-TH-2013-145, DCPT-13-86, IPPP-13-43, LPN13-036, MAN-HEP-2013-12",
    doi = "10.1007/JHEP09(2013)029",
    journal = "JHEP",
    volume = "09",
    pages = "029",
    year = "2013"
}

@article{ParticleDataGroup:2024cfk,
    author = "Navas, S. and others",
    collaboration = "Particle Data Group",
    title = "{Review of particle physics}",
    doi = "10.1103/PhysRevD.110.030001",
    journal = "Phys. Rev. D",
    volume = "110",
    number = "3",
    pages = "030001",
    year = "2024"
}

@article{Gallicchio:2012ez,
    author = "Gallicchio, Jason and Schwartz, Matthew D.",
    title = "{Quark and Gluon Jet Substructure}",
    eprint = "1211.7038",
    archivePrefix = "arXiv",
    primaryClass = "hep-ph",
    doi = "10.1007/JHEP04(2013)090",
    journal = "JHEP",
    volume = "04",
    pages = "090",
    year = "2013"
}

@article{Zaheer:2017wmg,
    author = "Zaheer, Manzil and Kottur, Satwik and Ravanbakhsh, Siamak and Poczos, Barnabas and Salakhutdinov, Ruslan and Smola, Alexander",
    title = "{Deep Sets}",
    eprint = "1703.06114",
    archivePrefix = "arXiv",
    primaryClass = "cs.LG",
    month = "3",
    year = "2017"
}

@article{Almeida:2008yp,
    author = "Almeida, Leandro G. and Lee, Seung J. and Perez, Gilad and Sterman, George F. and Sung, Ilmo and Virzi, Joseph",
    title = "{Substructure of high-$p_T$ Jets at the LHC}",
    eprint = "0807.0234",
    archivePrefix = "arXiv",
    primaryClass = "hep-ph",
    reportNumber = "YITP-SB-08-31",
    doi = "10.1103/PhysRevD.79.074017",
    journal = "Phys. Rev. D",
    volume = "79",
    pages = "074017",
    year = "2009"
}

@article{Yan:2020zrz,
    author = "Yan, Jun and Chen, Shi-Yong and Dai, Wei and Zhang, Ben-Wei and Wang, Enke",
    title = "{Medium modifications of girth distributions for inclusive jets and $Z^0+{\rm jet}$ in relativistic heavy-ion collisions at the LHC}",
    eprint = "2005.01093",
    archivePrefix = "arXiv",
    primaryClass = "hep-ph",
    doi = "10.1088/1674-1137/abca2b",
    journal = "Chin. Phys. C",
    volume = "45",
    number = "2",
    pages = "024102",
    year = "2021"
}

@article{Reichelt:2021svh,
    author = "Reichelt, Daniel and Caletti, Simone and Fedkevych, Oleh and Marzani, Simone and Schumann, Steffen and Soyez, Gregory",
    title = "{Phenomenology of jet angularities at the LHC}",
    eprint = "2112.09545",
    archivePrefix = "arXiv",
    primaryClass = "hep-ph",
    reportNumber = "MCNET-21-36, IPPP/21/57",
    doi = "10.1007/JHEP03(2022)131",
    journal = "JHEP",
    volume = "03",
    pages = "131",
    year = "2022"
}

@article{Krohn:2012fg,
    author = "Krohn, David and Schwartz, Matthew D. and Lin, Tongyan and Waalewijn, Wouter J.",
    title = "{Jet Charge at the LHC}",
    eprint = "1209.2421",
    archivePrefix = "arXiv",
    primaryClass = "hep-ph",
    doi = "10.1103/PhysRevLett.110.212001",
    journal = "Phys. Rev. Lett.",
    volume = "110",
    number = "21",
    pages = "212001",
    year = "2013"
}

@article{Kang:2023ptt,
    author = "Kang, Zhong-Bo and Larkoski, Andrew J. and Yang, Jinghong",
    title = "{Towards a Nonperturbative Formulation of the Jet Charge}",
    eprint = "2301.09649",
    archivePrefix = "arXiv",
    primaryClass = "hep-ph",
    doi = "10.1103/PhysRevLett.130.151901",
    journal = "Phys. Rev. Lett.",
    volume = "130",
    number = "15",
    pages = "151901",
    year = "2023"
}

@article{Berge:1980dx,
    author = "Berge, J. P. and others",
    title = "{Quark Jets from Antineutrino Interactions I: Net Charge and Factorization in the Quark Jets}",
    reportNumber = "FERMILAB-PUB-80-062-EXP, FERMILAB-PUB-80-062-E",
    doi = "10.1016/0550-3213(81)90207-8",
    journal = "Nucl. Phys. B",
    volume = "184",
    pages = "13--30",
    year = "1981"
}

@article{ALEPH:1991fba,
    author = "Decamp, D. and others",
    collaboration = "ALEPH",
    title = "{Measurement of charge asymmetry in hadronic Z decays}",
    reportNumber = "CERN-PPE-91-27, FSU-SCRI-91-45",
    doi = "10.1016/0370-2693(91)90844-G",
    journal = "Phys. Lett. B",
    volume = "259",
    pages = "377--388",
    year = "1991"
}

@article{Ngairangbam:2023cps,
    author = "Ngairangbam, Vishal S. and Spannowsky, Michael",
    title = "{Interpretable deep learning models for the inference and classification of LHC data}",
    eprint = "2312.12330",
    archivePrefix = "arXiv",
    primaryClass = "hep-ph",
    reportNumber = "IPPP/23/81",
    doi = "10.1007/JHEP05(2024)004",
    journal = "JHEP",
    volume = "05",
    pages = "004",
    year = "2024"
}

@article{taskinKnowledgeIntegrationPhysicsinformed2026,
    title = {Knowledge integration for physics-informed symbolic regression using pre-trained large language models},
    volume = {16},
    copyright = {2026 The Author(s)},
    issn = {2045-2322},
    url = {https://www.nature.com/articles/s41598-026-35327-6},
    doi = {10.1038/s41598-026-35327-6},
    language = {en},
    number = {1},
    urldate = {2026-06-04},
    journal = {Scientific Reports},
    publisher = {Nature Publishing Group},
    author = {Taskin, Bilge and Xie, Wenxiong and Lazebnik, Teddy},
    month = jan,
    year = {2026},
    pages = {1614},
}

@article{Bogatskiy:2023nnw,
    author = "Bogatskiy, Alexander and Hoffman, Timothy and Miller, David W. and Offermann, Jan T. and Liu, Xiaoyang",
    title = "{Explainable equivariant neural networks for particle physics: PELICAN}",
    eprint = "2307.16506",
    archivePrefix = "arXiv",
    primaryClass = "hep-ph",
    doi = "10.1007/JHEP03(2024)113",
    journal = "JHEP",
    volume = "03",
    pages = "113",
    year = "2024"
}

@article{Larkoski:2014pca,
    author = "Larkoski, Andrew J. and Thaler, Jesse and Waalewijn, Wouter J.",
    title = "{Gaining (Mutual) Information about Quark/Gluon Discrimination}",
    eprint = "1408.3122",
    archivePrefix = "arXiv",
    primaryClass = "hep-ph",
    reportNumber = "MIT--CTP-4572, NIKHEF-2014-026",
    doi = "10.1007/JHEP11(2014)129",
    journal = "JHEP",
    volume = "11",
    pages = "129",
    year = "2014"
}

@article{Berger:2003iw,
    author = "Berger, Carola F. and Kucs, Tibor and Sterman, George F.",
    title = "{Event shape / energy flow correlations}",
    eprint = "hep-ph/0303051",
    archivePrefix = "arXiv",
    reportNumber = "YITP-SB-03-06",
    doi = "10.1103/PhysRevD.68.014012",
    journal = "Phys. Rev. D",
    volume = "68",
    pages = "014012",
    year = "2003"
}

@article{Cho:2020ftg,
    author = "Cho, Won Sang and Kim, Hyung Do and Lee, Dongsub",
    title = "{Boosting invisible Higgs boson searches by tagging a gluon jet for the gluon fusion process}",
    eprint = "2003.06822",
    archivePrefix = "arXiv",
    primaryClass = "hep-ph",
    doi = "10.1103/PhysRevD.102.115007",
    journal = "Phys. Rev. D",
    volume = "102",
    number = "11",
    pages = "115007",
    year = "2020"
}

@article{ATLAS:2016fbc,
    author = "Aaboud, Morad and others",
    collaboration = "ATLAS",
    title = "{Measurement of the W boson polarisation in $t\bar{t}$ events from pp collisions at $\sqrt{s}$ = 8 TeV in the lepton + jets channel with ATLAS}",
    eprint = "1612.02577",
    archivePrefix = "arXiv",
    primaryClass = "hep-ex",
    reportNumber = "CERN-EP-2016-219, CERN-PH-2016-219",
    doi = "10.1140/epjc/s10052-017-4819-4",
    journal = "Eur. Phys. J. C",
    volume = "77",
    number = "4",
    pages = "264",
    year = "2017",
    note = "[Erratum: Eur.Phys.J.C 79, 19 (2019)]"
}

@article{CMS:2020ezf,
    author = "Aad, Georges and others",
    collaboration = "CMS, ATLAS",
    title = "{Combination of the W boson polarization measurements in top quark decays using ATLAS and CMS data at $\sqrt{s} =$ 8 TeV}",
    eprint = "2005.03799",
    archivePrefix = "arXiv",
    primaryClass = "hep-ex",
    reportNumber = "CMS-TOP-19-004, ATLAS-TOPQ-2018-02, CERN-EP-2020-012",
    doi = "10.1007/JHEP08(2020)051",
    journal = "JHEP",
    volume = "08",
    number = "08",
    pages = "051",
    year = "2020"
}

@article{Schuld:2020enb,
    author = "Schuld, Maria and Sweke, Ryan and Meyer, Johannes Jakob",
    title = "{Effect of data encoding on the expressive power of variational quantum-machine-learning models}",
    eprint = "2008.08605",
    archivePrefix = "arXiv",
    primaryClass = "quant-ph",
    doi = "10.1103/PhysRevA.103.032430",
    journal = "Phys. Rev. A",
    volume = "103",
    number = "3",
    pages = "032430",
    year = "2021"
}

@article{Shaw:2020udc,
    author = "Shaw, Alexander F. and Lougovski, Pavel and Stryker, Jesse R. and Wiebe, Nathan",
    title = "{Quantum Algorithms for Simulating the Lattice Schwinger Model}",
    eprint = "2002.11146",
    archivePrefix = "arXiv",
    primaryClass = "quant-ph",
    reportNumber = "INT-PUB-20-008",
    doi = "10.22331/q-2020-08-10-306",
    journal = "Quantum",
    volume = "4",
    pages = "306",
    year = "2020"
}

@article{Cogan:2014oua,
    author = "Cogan, Josh and Kagan, Michael and Strauss, Emanuel and Schwarztman, Ariel",
    title = "{Jet-Images: Computer Vision Inspired Techniques for Jet Tagging}",
    eprint = "1407.5675",
    archivePrefix = "arXiv",
    primaryClass = "hep-ph",
    doi = "10.1007/JHEP02(2015)118",
    journal = "JHEP",
    volume = "02",
    pages = "118",
    year = "2015"
}

@article{Davoudi:2024wyv,
    author = "Davoudi, Zohreh and Hsieh, Chung-Chun and Kadam, Saurabh V.",
    title = "{Scattering wave packets of hadrons in gauge theories: Preparation on a quantum computer}",
    eprint = "2402.00840",
    archivePrefix = "arXiv",
    primaryClass = "quant-ph",
    reportNumber = "UMD-PP-024-02, IQuS@UW-21-071",
    doi = "10.22331/q-2024-11-11-1520",
    journal = "Quantum",
    volume = "8",
    pages = "1520",
    year = "2024"
}

\end{document}